\documentclass[preprint,aps,prc,showpacs,nofootinbib]{revtex4-1}

\usepackage{graphicx}% Include figure files
\usepackage{bm}% bold math

\begin{document}

\title{Possibility of cold nuclear compression in antiproton-nucleus collisions}

\author{A.B. Larionov$^{1,2,3}$\footnote{Corresponding author.\\ 
        E-mail address: larionov@fias.uni-frankfurt.de}, 
        I.N. Mishustin$^{1,2}$, L.M. Satarov$^{1,2}$, 
        and W. Greiner$^1$}

\affiliation{$^1$Frankfurt Institute for Advanced Studies, J.W. Goethe-Universit\"at,
             D-60438 Frankfurt am Main, Germany\\ 
             $^2$Russian Research Center ``Kurchatov Institute'', 
             123182 Moscow, Russia\\
             $^3$Institut f\"ur Theoretische Physik, Universit\"at Giessen,
             D-35392 Giessen, Germany}

\date{\today}

\begin{abstract}
We study the dynamical response of the $^{16}$O nucleus to an incident
antiproton using the Giessen Boltzmann-Uehling-Uhlenbeck microscopic
transport model with relativistic mean fields. A special emphasis
is put on the possibility of a dynamical compression of the 
nucleus induced by the moving antiproton. Realistic antibaryon 
coupling constants to the mean meson fields are chosen in accordance
with empirical data. Our calculations show that an antiproton
embedded in the nuclear interior with momentum less than the nucleon Fermi
momentum may create a locally compressed zone in the nucleus with a maximum
density of about twice the nuclear saturation density.  
To evaluate the probability of the nuclear compression in high-energy
$\bar p$-nucleus collisions, we adopt a two-stage scheme.
This scheme takes into account the antiproton deceleration due to
the cascade of $\bar p N$ rescatterings inside the nucleus (first stage) 
as well as the nuclear compression by the slow antiproton before its annihilation
(second stage).
With our standard model parameters, the fraction of $\bar p$ annihilation events
in the compressed zone is about $10^{-5}$ for $\bar p ^{16}$O collisions at
$p_{\rm lab}=3-10$ GeV/c.
Finally, possible experimental triggers aimed at selecting such events 
are discussed.
\end{abstract} 

\pacs{25.43.+t;21.30.Fe;24.10.Lx}

\maketitle

\section{Introduction}
\label{Intro}

The production of compressed nuclear matter in laboratory is one of the most
important achievements of heavy-ion physics during last decades.
Heavy-ion collision experiments open the possibility to study new phases of matter, 
such as e.g. a quark-gluon plasma \cite{Collins:1974ky,Shuryak:1980tp}
(see also \cite{braun-munzinger:1031} for a recent review).
In a heavy-ion collision, compression is accompanied by the strong heating
of matter by the shock wave mechanism \cite{PhysRevLett.36.88}. 
However, very little is known about possible compressional effects induced by
a slowly moving or even stopped hadron in a nucleus. In this case,
the compression is associated with the enhanced concentration of nucleons
around the hadron, provided its interaction with nucleons is sufficiently 
attractive. Several examples of such systems are under discussion,
but their existence is still an open question.

The most famous example of strongly-bound hadron-nucleus systems are 
$\Lambda$-hypernuclei.
By measuring the $E2(5/2^+ \to 1/2^+)$ transition in the $^7_\Lambda$Li hypernucleus, 
the shrinkage of the $^6$Li core size by $\sim 19\%$
has been found experimentally in Ref. \cite{PhysRevLett.86.1982}.
The hypothetical multi-strange nuclei composed of several hyperons 
($\Lambda,\Xi^0,\Xi^-$) and nucleons (see \cite{Schaffner:1993nn,Schaffner199435}
and references therein) could be selfbound and have an enhanced baryon
density.

As proposed in Ref. \cite{Akaishi:2002bg} on the basis of a phenomenological
potential model, hypothetical $\bar K$-nuclei could be long-lived and strongly bound compact 
systems with the nucleon density reaching almost $10\rho_0$, where 
$\rho_0$ is the normal nuclear matter density.
However, as stated in the recent work by Hayano et al \cite{Hayano:2008zzd},
the measurement of 2p shift in the 3d-2p transition in kaonic 
$^4$He, eliminating a long-standing discrepancy between the standard 
theory and experiment, poses severe limitations on superstrong 
potentials with such high densities. 
Moderate compressional (core polarization) effects, up to about $2\rho_0$,
have been found within the relativistic mean field calculations of $\bar K$-nuclei 
done in Ref. \cite{Mares:2006vk}. 
However, the depths of 100-200 MeV for the real part of the $\bar K$-nucleus potential
at $\rho_0$, which follow from the phenomenological models of Refs. 
\cite{Akaishi:2002bg,Mares:2006vk}, are disfavoured by the chiral SU(3) models 
\cite{Ramos:1999ku,Oset:2005sn,Hyodo:2007jq}.
Nevertheless, the existence of strongly-bound $\bar K$-nuclear systems 
largely related to the nature of $\Lambda(1405)$ is still
under theoretical debates \cite{Dote:2008hw,Akaishi:2010wt}
and experimental studies \cite{Yamazaki:2010mu}.

Recently strong compressional effects have been predicted in the strongly
bound $\bar p$-nucleus systems \cite{Buervenich:2002ns,Mishustin:2004xa} in the case
of a deep real part of an antiproton optical potential, 
$\mbox{Re}(V_{\rm opt}) < -(150-200)$ MeV at $\rho_0$. 
We remark, however, that the antibaryon optical potentials
in the nuclear interior are largely unknown and their study requires more efforts
\cite{Friedman:2005ad,Pochodzalla:2008ju,PANDA,Larionov:2009tc,Hernandez:1986ev,
Hernandez:1989ij,Hernandez:1992rv}.

In the present work, we extend our previous study of the dynamical compression 
induced by a stopped antiproton \cite{Larionov:2008wy} to the case of 
a {\it moving} $\bar p$. The calculations are based on the Giessen 
Boltzmann-Uehling-Uhlenbeck (GiBUU) transport model \cite{GiBUU}. 
First, we study kinematical and geometrical conditions at which an 
antiproton can generate the increase of nucleon density.
Second, by performing the transport simulations of $\bar p$-nucleus
collisions we evaluate the actual probability of $\bar p$-annihilation 
in the compressed zone for the beam momenta of 0.3-10 GeV/c,
relevant for future antiproton beams at the Facility for Antiproton
and Ion Research (FAIR) in Darmstadt. Finally, we study possible
triggering schemes which can be used to select the events with 
$\bar p$-annihilation in the compressed nuclear environment. 
We have chosen the $^{16}$O nucleus as a target. This is
motivated by our earlier observation  
\cite{Buervenich:2002ns,Mishustin:2004xa,Larionov:2008wy}
that the compressional effects associated with $\bar p$ 
are more pronounced in light nuclei.

The paper is organized as follows: In Sect.~\ref{Procedure}, 
the description of a calculational procedure is given. 
Then, in Sect.~\ref{Dynamics}, we study the dynamical patterns 
of the nuclear compression by an antiproton initialized at 
different momenta and positions inside a nucleus. 
Sect.~\ref{Collisions} contains our results on the probabilities of a 
$\bar p$-annihilation in the compressed zone for energetic 
$\bar p$-nucleus collisions. In Sect.~\ref{Collisions}, we also discuss
possible triggers based on the fast proton emission and on the measurement 
of the energy deposition. We analyse the influence of the possible in-medium
modifications of the $\bar p$-annihilation rate and of the different antiproton
mean field parameters on our results in Sect.~\ref{Inmedium}.  
Summary and outlook are given in Sect.~\ref{Summary}.

\section{The calculational procedure}
\label{Procedure}

In our calculations, we apply the GiBUU model \cite{GiBUU}. This model solves
the coupled set of semiclassical kinetic equations for various hadronic
species: nucleons, antinucleons, mesons, baryonic resonances and their 
corresponding antiparticles. We use the relativistic mean field 
mode of calculations \cite{Larionov:2008wy,Larionov:2007hy,
Gaitanos:2007mm,Larionov:2009tc} which provides a simple and natural 
description of both baryonic and antibaryonic mean fields by using the same 
Lagrangian. The kinetic equation for the hadron of the type $j$ 
($j=p,~n,~\Delta^{++},~\Delta^+,~\Delta^0,~\Delta^-,...,~\pi^+,~\pi^0,~\pi^-,...$
and respective antibaryons) reads as
\begin{equation}
  \frac{1}{p_0^*}
  \left[ p^{*\mu} \frac{\partial}{\partial x^\mu} + \left(p_\mu^* F_j^{k\mu} 
                                   + m_j^* \frac{\partial m_j^*}{\partial x_k}\right)
    \frac{\partial}{\partial p^{* k}} \right] f_j(x,{\bf p^*}) 
  = I_j[\{f\}]~,                                                 \label{kinEq}
\end{equation}
where $f_j(x,{\bf p^*})$ is the phase-space density of the $j$-th type particles,
$p^*$ is the kinetic four-momentum ($p_0^*=\sqrt{ {\bf p^*}^2 + (m_j^*)^2 }$), 
$m_j^*$ is the effective mass, and $F_j^{k\mu}$ is the field tensor.
The l.h.s. of Eq. (\ref{kinEq}) describes the evolution of the phase-space density 
$f_j(x,{\bf p^*})$ under the influence of the mean mesonic fields. The r.h.s. of Eq. (\ref{kinEq})
is the collision integral $I_j[\{f\}]$ describing the change of the phase-space density due to 
the particle-particle collisions and resonance decays. 

The kinetic equations (\ref{kinEq}) are solved by applying the standard test particle technique
in the parallel ensemble mode. The phase-space densities are represented by the set of the 
point-like test particles:
\begin{equation}
   f_j(x,{\bf p}^*) = \frac{ (2\pi)^3 }{ g_j N_{\rm ens} } 
                      \sum_{i=1}^{ N_{\rm ens} N_j} 
                      \delta( {\bf r} - {\bf r}_i(t) )\,
                      \delta( {\bf p}^* - {\bf p}^*_i(t) )~,      \label{tP}
\end{equation}
where $N_j$ is the number of physical particles of the type $j$, $N_{\rm ens}$ is the
number of parallel ensembles, and $g_j$ is the spin degeneracy. 
The test-particle representation (\ref{tP}) provides a simple solution of 
the kinetic equations (\ref{kinEq}) in terms of the Hamiltonian-like
equations for the centroids $({\bf r}_i(t),{\bf p}^*_i(t))$ (c.f. Eqs.(2),(3)
in Ref. \cite{Larionov:2009tc}). The collision integral is simulated with
the help of a usual geometrical collision criterion (c.f. Ref. \cite{Larionov:2007hy}).

The mean mesonic fields are determined from the nonlinear Klein-Gordon-like equations
with the source terms given by the particle densities and currents.
Therefore, in order to provide a smooth coordinate dependence of the mean mesonic fields,
the coordinate space $\delta$-functions in the r.h.s. of Eq. (\ref{tP}) are replaced by 
the Gaussians of the width $L \simeq 0.5-1$ fm in actual calculations. 
Then, e.g. the coordinate space density and the scalar density of the $j$-th 
type hadrons are computed as
\begin{eqnarray}
   \rho_j(x) &=& \frac{g_j}{(2\pi)^3} \int\,d^3 p^\star
               f_j(x,{\bf p^\star})
              = \frac{ 1 }{ (2\pi)^{3/2} L^3 N_{\rm ens} } 
                \sum_{i=1}^{ N_{\rm ens} N_j}
                \exp\left\{-\frac{({\bf r} - {\bf r}_i(t))^2}{2L^2}\right\} 
                              ~,                            \label{rho_j}\\
   \rho_{Sj}(x) &=& \frac{g_j}{(2\pi)^3} \int\,d^3 p^\star
               \frac{m_j^*}{p^{*0}}
               f_j(x,{\bf p^\star})
              = \frac{ 1 }{ (2\pi)^{3/2} L^3 N_{\rm ens} } 
                \sum_{i=1}^{ N_{\rm ens} N_j }
                \frac{m_i^*}{p^{*0}_i}
                \exp\left\{-\frac{({\bf r} - {\bf r}_i(t))^2}{2L^2}\right\} 
                                  ~.                       \label{rho_Sj}
\end{eqnarray}
We are interested, in particular,  in the values of the nucleon density 
$\rho=\rho_p+\rho_n$. The antiproton density $\rho_{\bar p}$ and the nucleon scalar density 
$\rho_S=\rho_{Sp}+\rho_{Sn}$ are also used in the present analysis.

The width $L$ in Eqs.(\ref{rho_j}),(\ref{rho_Sj}) is a pure numerical parameter 
of the GiBUU model. Its value is correlated with the number of parallel ensembles 
and is set equal to the coordinate grid step size (c.f. \cite{Larionov:2007hy,
Larionov:2008wy}). The physical results do not depend on $L$, provided that it is 
small enough to resolve the physical nonuniformities of the system. In the present 
calculations we use the value $L=0.5$ fm from our earlier work \cite{Larionov:2008wy},
where we have also studied the influence of $L$ on the compression dynamics. 

For the nucleon mean field we apply the nonlinear Walecka model.  
The nucleon-meson coupling constants and the parameters of the 
$\sigma$-field self-interactions are taken from the NL3 parameterization 
\cite{Lalazissis:1996rd}.
This parameterization provides the nuclear matter incompressibility 
$K=272$ MeV and the nucleon effective mass $m_N^*=0.6m_N$ at $\rho_0=0.148$ 
fm$^{-3}$.
Within the NL3 set of parameters,
the binding energies, charge and neutron radii of spherical nuclei
as well as deformation properties of some rare-earth and actinide nuclei 
have been described quite well \cite{Lalazissis:1996rd}. The isoscalar monopole 
resonance energies in heavy spherical nuclei are also reproduced by this 
set of parameters \cite{Lalazissis:1996rd}.

The antinucleon-meson coupling constants are more uncertain. 
As it is well known, the G-parity transformation of Walecka-type Lagrangians 
results in too deep antiproton optical potentials. Therefore, following Refs. 
\cite{Mishustin:2004xa,Friedman:2005ad,Larionov:2008wy,Larionov:2009tc},
we introduce a common reduction factor $\xi < 1$ for the antinucleon 
coupling constants to the $\sigma$-, $\omega$- and $\rho$-mesons as given 
by the G-parity transformation. Below, if it is not explicitly stated otherwise, 
we use the value $\xi=0.22$ obtained in \cite{Larionov:2009tc} from the best fit of
$\bar p$-absorption cross sections on nuclei at the beam momenta below 1 GeV/c.
The corresponding real part of an antiproton optical potential is 
about -150 MeV in the nuclear centre, which is somewhat deeper than the real part 
derived from the most recent $\bar p$-atomic calculations \cite{Friedman:2005ad},
however, within the commonly accepted uncertainty interval%
\footnote{For detailed discussion, see \cite{Larionov:2009tc} and references therein}.  

Due to a big annihilation cross section, in majority of events, an antiproton 
colliding with a nucleus will annihilate already on peripheral nucleons.
However, as argued in Ref. \cite{Mishustin:2004xa}, compressional effects are
expected only in events when the antiproton penetrates deep to the nuclear interior
and stops there due to (in)elastic collisions with nucleons. Such events
are presumably quite rare and their study requires to go beyond the
ensemble-averaged description provided by the kinetic mean field theory.
The Quantum Molecular Dynamics \cite{Aichelin:1991xy} or Antisymmetrized 
Molecular Dynamics \cite{Dote:2005nb} models seem to be better theoretical 
tools for studying such rare events. However, to our knowledge, at present 
there exists no version of a molecular dynamics model which incorporates all 
relevant antibaryon-baryon collision channels and relativistic potentials.

In the present work, we treat compressional effects in a $\bar p$-nucleus 
collision perturbatively. It means, that the influence of the 
compressional response of a nucleus on the deceleration process and
eventual annihilation of an antiproton is neglected. Thus,
the collisional dynamics of the incident antiproton is simulated
within standard GiBUU until its annihilation. We assume further, that
the position and momentum of the $\bar p$ at the beginning of compression
process are not much different from those at its annihilation point. 
Then we study the compressional response of the nucleus to slow antiprotons
and evaluate their survival probability.

Therefore, we adopt a two-stage calculational scheme:
\underline{On the first stage}, an antiproton penetrates into the nucleus
while experiencing one or more rescatterings on nucleons. We describe
this process by the standard GiBUU simulation in the parallel ensemble
mode with $N_{\rm ens}=1000$ parallel ensembles. Each parallel ensemble is
considered as one event. For each impact parameter, $N_{\rm runs}=100$ simulation
runs have been done which gives $N_{\rm ens} N_{\rm runs} =10^5$
events per impact parameter. We have chosen 32 impact parameters
$b=0.25,~0.50,...,8$ fm for the $\bar p^{16}$O system. Final results
are impact-parameter weighted. 
Since the incoming $\bar p$ can be transformed to another
antibaryons, e.g. $\bar n$ or $\bar \Delta$, we consider below 
the antibaryon annihilation in general. The coordinates ${\bf r}_{\bar B}$ 
and the kinetic three-momenta ${\bf p}_{\bar B}^*$ of an antibaryon just 
before the annihilation or, for events without annihilation, at the end of 
the computational time (40 fm/c) have been determined and stored for every event. 
In the following, we always deal with the kinetic three-momenta of particles,
but omit the word ``kinetic'' and the star symbol for brevity%
\footnote{In fact, if the collective motion of nuclear matter 
is negligible, e.g. when a fast hadron passes through the undisturbed
nuclear target, the space components of the canonical and kinetic 
four-momenta are practically the same.}.
Due to the averaging of the mean field over parallel ensembles, the compressional 
effects are practically unnoticeable in the standard GiBUU calculation,
because rare events with a deep penetration of $\bar p$ into the nucleus
are diluted with the majority of events when the antiproton annihilates on the nuclear
periphery. This is why we use the coordinates and momenta of the antibaryon 
obtained on the first stage as an input for another simulation based on the GiBUU model
\cite{Larionov:2008wy,Mishustin:2008ch}. Thus, \underline{on the second stage}, 
an antiproton is initialized inside the nucleus at the phase-space point
$({\bf r}_{\bar B},{\bf p}_{\bar B})$ using the Gaussian distribution in coordinate space
and the sharp-peaked distribution in momentum space.
By doing so we neglect the possibility that the annihilating antibaryon can be different
from the antiproton. This is, however, not important in view that the mean field contributions,
apart from small isospin and Coulomb effects, are the same for all antibaryons in our model.
The corresponding phase-space density of an antiproton is written as $(\hbar=c=1)$: 
\begin{equation}
   f_{\bar p}({\bf r},{\bf p}) = \frac{1}{(2\pi)^{3/2}\sigma_r^3} 
                                 \exp\left\{ -\frac{({\bf r}-{\bf r}_{\bar B})^2}{2\sigma_r^2}\right\}
                                 \frac{ (2\pi)^3 }{ g_{\bar p} }
                                 \delta({\bf p}-{\bf p}_{\bar B})~,     \label{pbar_df}
\end{equation}
where $g_{\bar p}=2$ is the spin degeneracy of an antiproton and $\sigma_r$ 
is the width of the coordinate space Gaussian. Please, notice, that the quantity 
$\sigma_r$ is a physical parameter of our model, while the quantity $L$ in 
Eqs. (\ref{rho_j}),(\ref{rho_Sj}) is pure technical and should not be misidentified
with $\sigma_r$. We stress that now the antiproton test particles of all $N_{\rm ens}$ 
parallel ensembles are initialized according to 
Eq.~(\ref{pbar_df}) with the same centroid $({\bf r}_{\bar B},{\bf p}_{\bar B})$,
and the calculation is repeated for every event of the first stage.
Thus, the antiproton test particle contributions to the mean mesonic fields reflect 
the presence of a real antiproton at the phase-space point $({\bf r}_{\bar B},{\bf p}_{\bar B})$.
In this new calculation, therefore, the compressional effects will manifest themselves in 
full strength without dilution. Further evolution of the $\bar p$-nucleus system is
calculated in a similar way as in \cite{Larionov:2008wy} by using the GiBUU model
without annihilation.
However, in distinction to \cite{Larionov:2008wy}, we now take into account all 
collisional channels different from the annihilation one, in particular,
$NN \to NN$ and $\bar N N \to \bar N N$. This models dissipation leading to some
small heating of the nuclear system and slowing down the antiproton during 
compression process. For brevity, sometimes we refer to the GiBUU calculations
without annihilation as ``coherent'' calculations below.

Instead of explicitly treating the annihilation on the second stage of calculations,
we compute the survival probability of an antiproton in the course of compression as
\begin{equation}
   P_{\rm surv}(t) = \exp\left\{-\int\limits_0^t\,dt^\prime\,\Gamma_{\rm ann}(t^\prime)\right\}~.
                                               \label{Psurv}
\end{equation}
Here 
\begin{equation}
   \Gamma_{\rm ann}= < v_{\rm rel} \sigma_{\rm ann} > \rho   \label{Gamma_ann}
\end{equation}
is the antiproton width with respect to the annihilation, $\rho$ is the local nucleon density,
$v_{\rm rel}$ is the relative velocity of an antiproton and a nucleon and $\sigma_{\rm ann}$
is the $\bar p$-annihilation cross section on a nucleon. Brackets in Eq.~(\ref{Gamma_ann})
denote averaging over the nucleon Fermi motion. 

The two-stage scheme described above is not fully equivalent to the true molecular dynamics
simulation. However, the most interesting phenomenon which we want to study, i.e. 
the dynamical compression of a nucleus by a slow antiproton, can be realistically
simulated in this way.

As we will see below (c.f. Figs.~\ref{fig:rhoz} and \ref{fig:sigCompr}), the width $\sigma_r$
of the Gaussian in Eq.~(\ref{pbar_df}) is a very important parameter, which can not be determined
from our model. We, therefore, consider two most representative values: $\sigma_r=1$ fm and 
$\sigma_r=0.14$ fm. The first choice corresponds to a rather wide wave packet which presumably
describes the static wave function of a strongly bound antiproton
implanted in a nucleus \cite{Buervenich:2002ns,Mishustin:2004xa,Larionov:2008wy}. The second
choice of a narrow wave packet is adjusted to describe the charge r.m.s. radius of a physical
(anti)proton, $r_p=0.9$ fm \cite{Montanet:1994xu}. Indeed, in our model, the true source charge
distribution of an antiproton is given by folding the coordinate space Gaussian (\ref{pbar_df})
with the test particle Gaussian. Thus, we have $\sigma_p=\sqrt{\sigma_r^2+L^2}$, where 
$\sigma_p=r_p/\sqrt{3}$ is the charge distribution width of a physical (anti)proton.

The second-stage calculations can be significantly accelerated if one neglects the changes
in a target nucleus caused by the antiproton cascade on the first stage. Then, the spherical
symmetry of the $^{16}$O target nucleus can be utilized. In this case, the compressional
evolution depends only on three variables: the absolute values of the antiproton 
initial radius-vector ${\bf r}$ and momentum ${\bf p}$, and on the angle 
$\Theta=\arccos({\bf r p}/rp)$ between ${\bf r}$ and ${\bf p}$.
(One needs six variables ${\bf r}$, ${\bf p}$ in the case of arbitrary shape).
Therefore, the second-stage calculations have been performed with the target
nucleus for the set of the antiproton initial positions ${\bf r}$
and momenta ${\bf p}$ taken on the uniform $7\times20\times9$ grid in the space 
$(r,p,\cos\,\Theta)$, where $r\in[0.5;3.5]$ fm, $p\in[0.05;1.00]$ GeV/c,
and $\cos\,\Theta\in[-1;1]$.
The results of the second-stage calculations, in particular, the antiproton survival 
probabilities at the time moments corresponding to the system entering to
and exiting from the compressed state, have been stored. To determine the
compression probability for a given first-stage event, resulting coordinates
and momenta of the antibaryon at the annihilation point have been projected 
on the grid.

\section{Dynamics of nuclear compression}
\label{Dynamics}

In this section, the nuclear response to the moving antiproton is considered 
disregarding $\bar p$-annihilation. The latter is, however, implicitly taken
into account by following the time dependence of the $\bar p$-survival probability.   
\begin{figure}
\includegraphics[scale = 0.80]{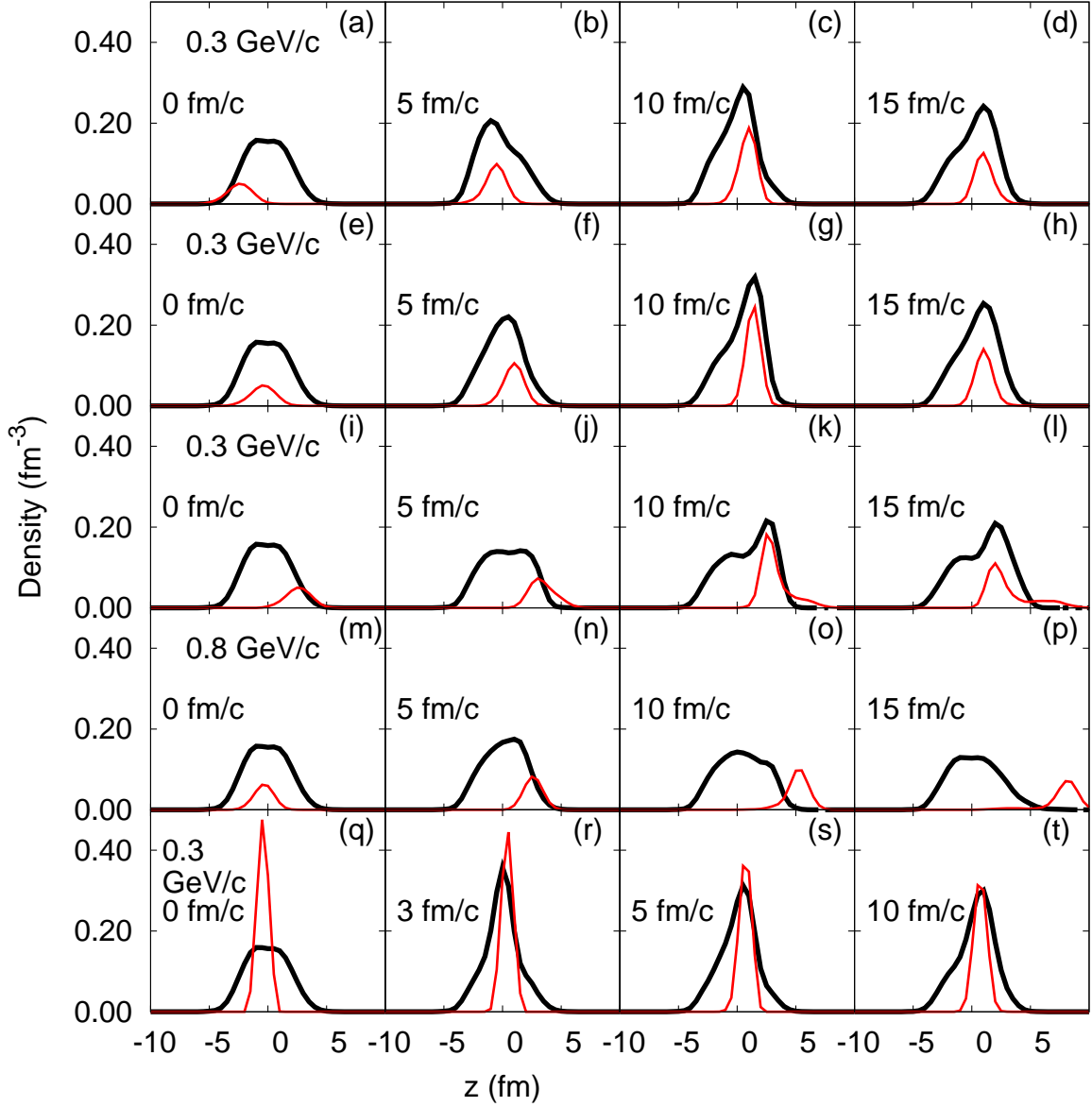}

\vspace*{0.5cm}

\caption{\label{fig:rhoz}(Color online) Nucleon (thick solid lines) and antiproton (thin solid lines)
densities as functions of the longitudinal coordinate $z$ at different time moments for 
the $\bar p^{16}$O system. The antiproton has been initialized on the axis passing through 
the nuclear centre, i.e. $x=y=0$ with momentum $p$ along the positive 
$z$-direction. Different rows correspond to different $\bar p$ initializations characterized
by the Gaussian width $\sigma_r$ (fm), momentum $p$ (GeV/c), and coordinate $z$ (fm): 
(a)-(d) --- $(\sigma_r,~p,~z)=$(1,~0.3,~-2.5);
(e)-(h) --- (1,~0.3,~-0.5); (i)-(l) --- (1,~0.3,~2.5);  
(m)-(p) --- (1,~0.8,~-0.5); (q)-(t) --- (0.14,~0.3,~-0.5).
The antibaryon annihilation is switched off in this calculation.}
\end{figure}

Figure~\ref{fig:rhoz} shows the time evolution of the nucleon and 
antiproton density distributions for the $\bar p ^{16}$O system for different 
$\bar p$ initializations. Only the cases are presented, when the initial 
antiproton momentum ${\bf p}$ is (anti)parallel to the initial position 
vector ${\bf r}$, i.e. $\Theta=0$ and $\Theta=\pi$. 

Let us start by considering how compression depends on the initial antiproton coordinate $z$ at fixed
momentum $p=0.3$ GeV/c. If the antiproton moves towards the nuclear centre, i.e. $\Theta=\pi$,
the compression of a nuclear system up to densities $\sim2\rho_0$ is reached within the time interval
of about $10$ fm/c (see panels (a)-(d) and (e)-(h) of  Fig.~\ref{fig:rhoz}).
For the outgoing antiproton ($\Theta=0$), the compression 
is much smaller (panels (i)-(l) of Fig.~\ref{fig:rhoz}), since the antiproton moves through
the nuclear periphery. It is interesting, that at $p=0.3$ GeV/c the antiproton does not
leave the nucleus but only bounces off the nuclear boundary and finally gets captured.
However, the capture takes place on the time scale of $\sim20$ fm/c and, therefore, would hardly
be observed due to a very low survival probability of the antiproton (see Fig.~\ref{fig:rhomax} below).

For a higher momentum $p=0.8$ GeV/c (panels (m)-(p) of Fig.~\ref{fig:rhoz}), the compression
is practically absent, since the nuclear response is much slower then the time needed by the antiproton
to cross the nucleus. We also see, that at $p=0.8$ GeV/c the antiproton escapes
from the nucleus, because its total in-medium energy 
$E_{\bar p}=\sqrt{{\bf p}^2+m_{\bar p}^{*\,2}}+V_{\bar p}^0$ exceeds the vacuum
mass $m_N$ by about 165 MeV. Here $m_{\bar p}^*= m_N + \xi (m_N^* - m_N)\simeq0.91m_N$
is the antiproton effective mass and $V_{\bar p}^0=-(308\,\xi)~\mbox{MeV}\simeq-68$ MeV
is the antiproton vector potential at $\rho=\rho_0$ 

The compression process is quite sensitive to the choice of initial Gaussian width of
the antiproton. One can see this from Fig.~\ref{fig:rhoz} by comparing the panels (e)-(h) and
(q)-(t), where the calculations are shown for the same initial positions and momenta 
of $\bar p$, but for the different widths $\sigma_r$.
Due to a deeper nucleon potential, a smaller initial $\bar p$-width
makes the compression more pronounced and fast.
Unless stated otherwise, the case of $\sigma_r=1$ fm is discussed below.

\begin{figure}
\includegraphics[scale = 0.80]{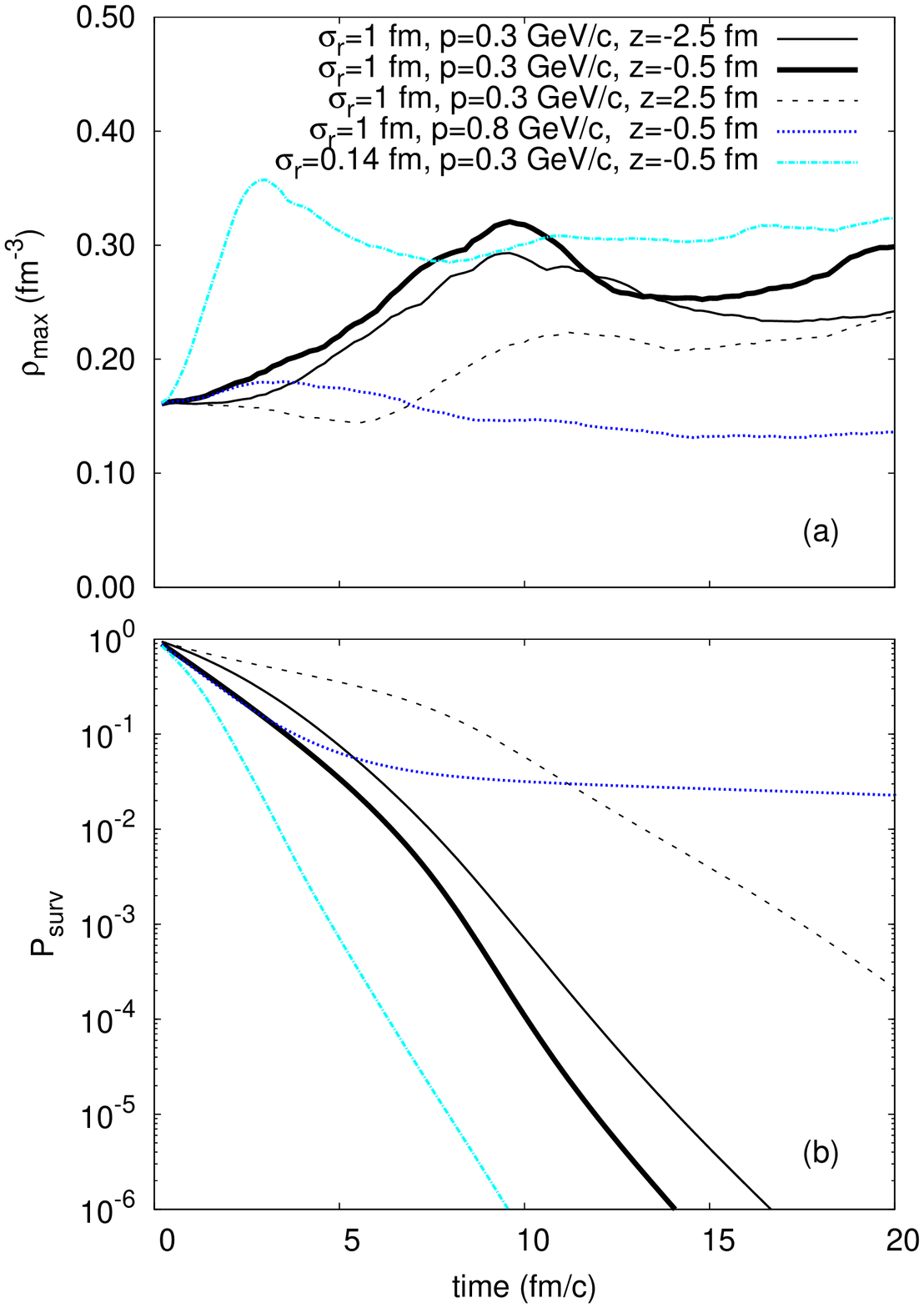}

\vspace*{1.5cm}

\caption{\label{fig:rhomax}(Color online) Maximum nucleon density (a) and antiproton survival
probability (b) as functions of time for the $\bar p^{16}$O system. Different curves correspond
to different $\bar p$-initializations as explained in Fig.~\ref{fig:rhoz}.}
\end{figure}

In Fig.~\ref{fig:rhomax}, we present the time dependence of the maximum nucleon density  
$\rho_{\rm max}$ and of the antiproton survival probability (\ref{Psurv}) for various 
antiproton initializations shown in Fig.~\ref{fig:rhoz}. As we have already seen in 
Fig.~\ref{fig:rhoz}, for the initializations with $z < 0$ (i.e. for $\Theta = \pi$)
and $p=0.3$ GeV/c, nucleon densities up to $2\rho_0$ are reached. The antiproton survives
with the probability $P_{\rm surv} \sim 10^{-2}$ until the time moment when the maximum
density $\rho_{\rm max}=2\rho_0$ is achieved.

The nuclear compression caused by an antiproton could only be observed, if the antiproton
would annihilate in the compressed nuclear environment. This process can be detected by
its specific final state characteristics. As shown in \cite{Mishustin:2004xa,Larionov:2008wy},
possible observable signals include the enhanced radial collective flow 
of nuclear fragments, hardening the energy spectra of emitted nucleons, and softening
the meson invariant mass distributions. Moreover, the multi-nucleon annihilation (MNA) 
channels with the baryonic number $B\geq1$ might be enhanced if the compressed zone is formed. 
A more exotic scenario, the deconfinement of an annihilation zone leading to the enhanced 
strangeness production has also being discussed in literature 
\cite{Mishustin:2004xa,Rafelski:1979nt,Salvini:2005sb, Bendiscioli:2009zza}.
Herein, we do not consider any specific signals caused by annihilation
in the compressed nuclear state. We rather concentrate on the evaluation of the total 
$\bar p$-annihilation probability at enhanced nucleon densities. For brevity, we refer
to this possibility as to the annihilation in a compressed zone (ACZ) below.

Let us define the compressed nuclear system as a system where the maximum nucleon 
density $\rho_{\rm max}$ exceeds some critical value $\rho_c$. 
If not stated otherwise, we choose $\rho_c=2\rho_0$ in calculations. 
Such density values can be reached, e.g. in central heavy-ion collisions
at beam energies of hundreds MeV/nucleon \cite{Stoecker:1986ci}. The probability
for the antiproton to annihilate at $\rho_{\rm max} > \rho_c$ is defined as 
\begin{equation}
   P_{\rm ann}^{\,\rm c} = P_{\rm surv}(t_1) - P_{\rm surv}(t_2)~,       \label{P_ann^c}
\end{equation}
where the time interval $[t_1;t_2]$ encloses the high-density phase of 
the time evolution, i.e. \mbox{$\rho_{\rm max}(t_1)=\rho_{\rm max}(t_2)=\rho_c$} with 
$\rho_{\rm max}(t) > \rho_c$ for $t_1 < t < t_2$
\footnote{When there are more than one such intervals, the earliest one is chosen.}.
For example, in the case $(\sigma_r,~p,~z)=$(1 fm, 0.3 GeV/c, -0.5 fm) we obtain
$t_1=8.4$ fm/c and $t_2=11$ fm/c (see Fig.~\ref{fig:rhomax}). Since the $\bar p$ 
survival probability drops exponentially with time, we have 
$P_{\rm surv}(t_1) \gg P_{\rm surv}(t_2)$ and, therefore, actually 
$P_{\rm ann}^{\,\rm c} \simeq P_{\rm surv}(t_1)$.

\begin{figure}
\includegraphics[scale = 0.80]{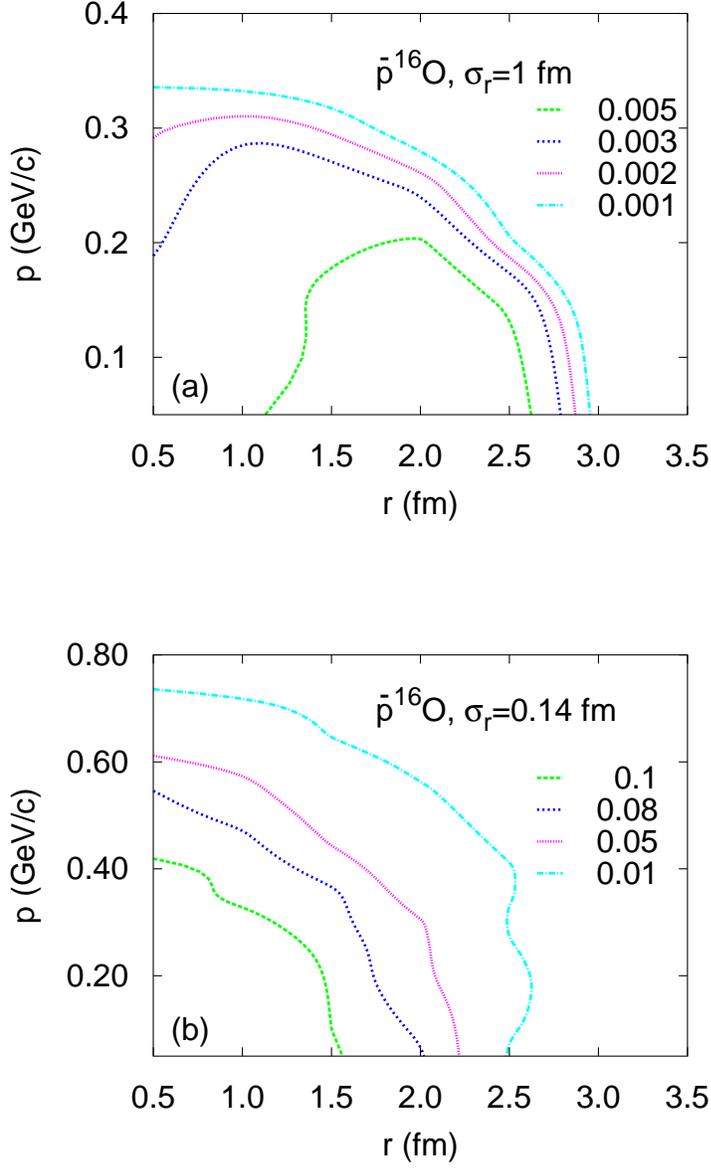}

\vspace*{0.5cm}

\caption{\label{fig:cont_psurv}(Color online)  The contour plots of the antiproton 
ACZ probability $P_{\rm ann}^{\,\rm c}$ at  $\rho_{\rm max} > 2\rho_0$ (see Eq.~(\ref{P_ann^c}))
in the plane given by initial values of the antiproton radial position $r$ and momentum $p$
for the system $\bar p^{16}$O. The values of $P_{\rm ann}^{\,\rm c}$ are averaged over 
the cosine of the angle between the initial radius vector and momentum of the antiproton. 
Panel (a)((b)) corresponds to the initial antiproton width $\sigma_r=1~(0.14)$ fm.}
\end{figure}
Figure~\ref{fig:cont_psurv} shows the antiproton ACZ probability
as a function of the $\bar p$ initial radial position and momentum. 
As expected, the $\bar p$-initializations with smaller momentum lead to larger 
$P_{\rm ann}^{\,\rm c}$. The radial dependence of $P_{\rm ann}^{\,\rm c}$ at fixed initial 
momentum is somewhat more complicated. In the case of a larger width of the 
initial antiproton space distribution ($\sigma_r=1$ fm), $P_{\rm ann}^{\,\rm c}$ 
has a weak maximum at $r \simeq 1-2$ fm and decreases towards the nuclear centre 
slightly.
This can be traced back to Fig.~\ref{fig:rhoz}, where we see, that the $\bar p$ 
initializations at different positions result in practically the same compressional 
effect provided that the antiproton moves to the nuclear centre (c.f. panels 
(a)-(d) and (e)-(h)). 
For a narrower initial antiproton space distribution ($\sigma_r=0.14$ fm), the
maximum of the ACZ probability is located at the nuclear centre, since compression
is much faster in this case, and, thus, is more sensitive to the local nucleon density.

\section{$\bar p$-nucleus collisions}
\label{Collisions}

As it was demonstrated in the previous section (c.f. Fig.~\ref{fig:cont_psurv}),
the ACZ probability depends on the position and momentum of the antiproton at 
the beginning of compression process.
Therefore, before discussing the results of a full two-stage calculation,
it is instructive to study the distributions of antibaryon annihilation points
in the coordinate and momentum space. These distributions are determined at the first
stage of calculations.

\begin{figure}
\includegraphics[scale = 0.80]{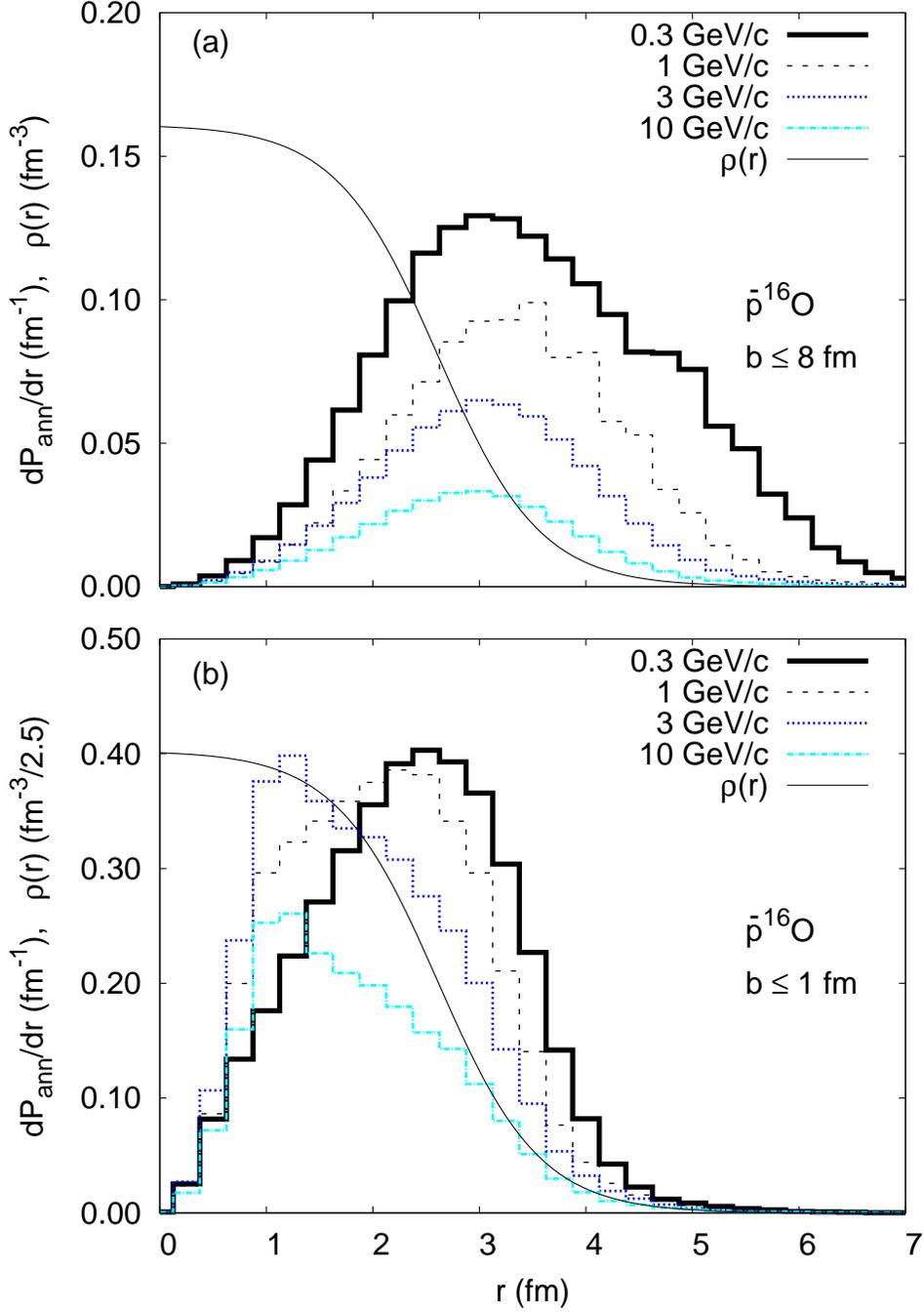}

\vspace*{0.5cm}

\caption{\label{fig:dsigdr}(Color online) Radial distributions of 
annihilation points for $\bar p^{16}$O collisions at different beam momenta
(see key notations) normalized to the total annihilation probability.
For the reference, thin solid lines show the nucleon density
profile of the $^{16}$O nucleus. Panels (a) and (b) represent the inclusive 
spectra ($b \leq 8$ fm) and the spectra for central collisions 
($b \leq 1$ fm), respectively. Note different scales of vertical axes
in panels (a) and (b).}
\end{figure}
Figure~\ref{fig:dsigdr} shows the radial distributions of the antibaryon 
annihilation points for the $\bar p^{16}$O reaction at several beam momenta.
For inclusive events (a), the maxima of these distributions are 
located at the peripheral region, where the density is about $30\%$ of
the central density, independent of the beam momentum. 
This is a pure geometrical effect caused by mixing of events
with all possible impact parameters. However, for central collisions (b),
the maxima are shifted closer to the nuclear centre. The shift becomes 
larger at higher beam momenta. This is expected, since with increasing 
$p_{\rm lab}$ the $\bar p$-nucleon annihilation cross section drops 
\cite{Cugnon:1989} leading to the larger fraction of deeply-located 
annihilations.

\begin{figure}
\includegraphics[scale = 0.80]{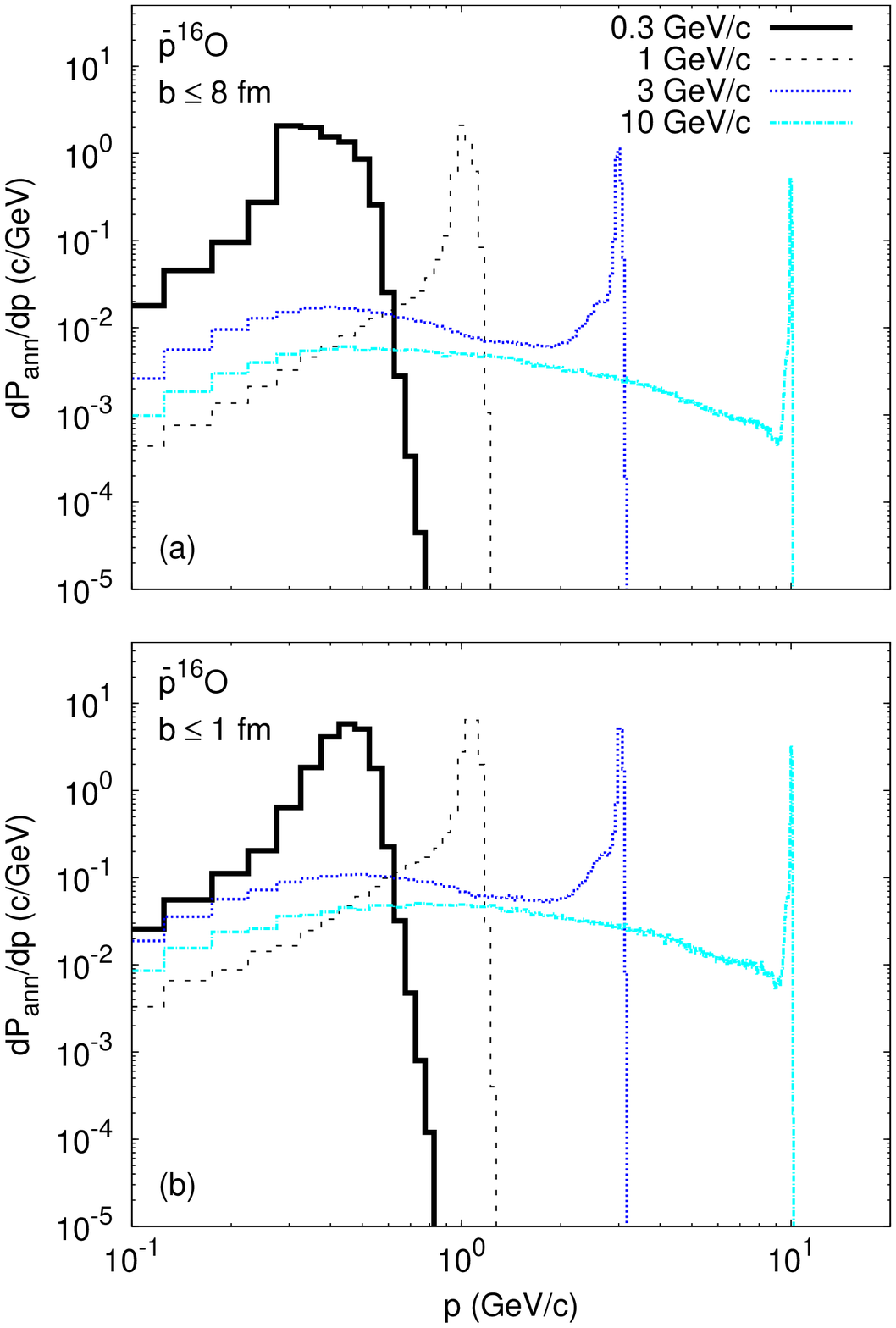}

\vspace*{0.5cm}

\caption{\label{fig:dsigdp}(Color online) Momentum distributions of 
annihilation points for $\bar p^{16}$O collisions at different beam momenta
(see key notations). Panels (a) and (b) represent the inclusive 
spectra and the spectra for the central collisions, respectively.}
\end{figure}
Figure~\ref{fig:dsigdp} demonstrates the momentum distributions of 
antibaryons at the annihilation points. There is only a little difference 
between the shapes of the distributions for the inclusive (a) and central
(b) events at the same beam momentum. However, the total annihilation
probability is increased by a factor of 3-10 for the central collisions,
which is also seen in Fig.~\ref{fig:dsigdr}. The distributions have
a sharp peak at the beam momentum and a long tail towards 
small momenta. In the case of the smallest beam momentum 
$p_{\rm lab}=0.3$ GeV/c, the peak is broader and shifted to the higher momenta 
$p > p_{\rm lab}$ for the central collisions. This is caused by the antiproton 
elastic collisions with the Fermi sea nucleons and by the antibaryon
acceleration in a strongly attractive mean field potential.

\begin{figure}
\includegraphics[scale = 0.80]{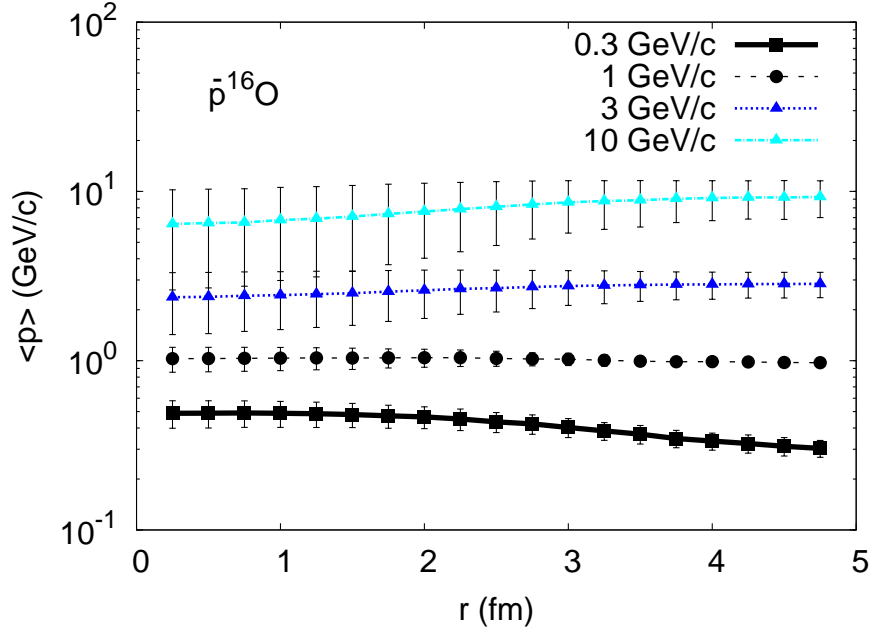}

\vspace*{0.5cm}

\caption{\label{fig:p_vs_r}(Color online) Average momentum of annihilating
baryon as a function of the radial position for the inclusive set of events
$\bar p^{16}$O at different beam momenta. The error bars represent the 
dispersion of momentum distribution at a given $r$.}
\end{figure}
The acceleration is better visible in Fig.~\ref{fig:p_vs_r} which shows 
the correlation between the radial position and momentum of an antibaryon
at the annihilation point.
The centrality dependence is quite weak in this case, thus we have presented
the results for the inclusive event set only. For $p_{\rm lab}=0.3$ GeV/c, the
average momentum of annihilating antibaryon increases up to 0.5 GeV/c at the
nuclear centre. For larger beam momenta, the mean field acceleration is hindered
by the collisional damping of an initial $\bar p$ momentum.  

We will now discuss the results of the full two-stage calculations (see 
Sect.~\ref{Procedure}).
The total annihilation cross section on a nucleus $\sigma_{\rm ann}$ and the ACZ
cross section $\sigma_{\rm compr}$ are determined as follows:
\begin{eqnarray}
&& \sigma_{\rm ann} = \sum_{b \leq b_{\rm max}} 2\pi b \Delta b 
                      \frac{N_{\rm ann}(b)}{N_{\rm ev}(b)}~, \label{sigma_ann} \\
&& \sigma_{\rm compr}= \sum_{b \leq b_{\rm max}} 2\pi b \Delta b 
                       \frac{1}{N_{\rm ev}(b)}
                       \sum_{i=1}^{N_{\rm ann}(b)} 
                       P_{\rm ann}^{\,\rm c}({\bf r}_i,{\bf p}_i)~. \label{sigma_compr}
\end{eqnarray}   
Here, $N_{\rm ev}(b)$ and $N_{\rm ann}(b)$ are, respectively, the total number 
of events and the number of annihilation events calculated within standard
GiBUU (the first stage) for a given impact parameter $b$.
The quantity $P_{\rm ann}^{\,\rm c}({\bf r}_i,{\bf p}_i)$ (see Eq.~(\ref{P_ann^c})),
which depends on the annihilation point position ${\bf r}_i$ and momentum ${\bf p}_i$
in a given annihilation event $i$, is the annihilation probability at 
$\rho_{\rm max} > \rho_c$ computed within a coherent GiBUU run (the second stage). 
The cutoff value of the impact parameter $b_{\rm max}$ has been
chosen to be 8 fm for an inclusive event set and 1 fm for central events. 
 
\begin{figure}
\includegraphics[scale = 0.80]{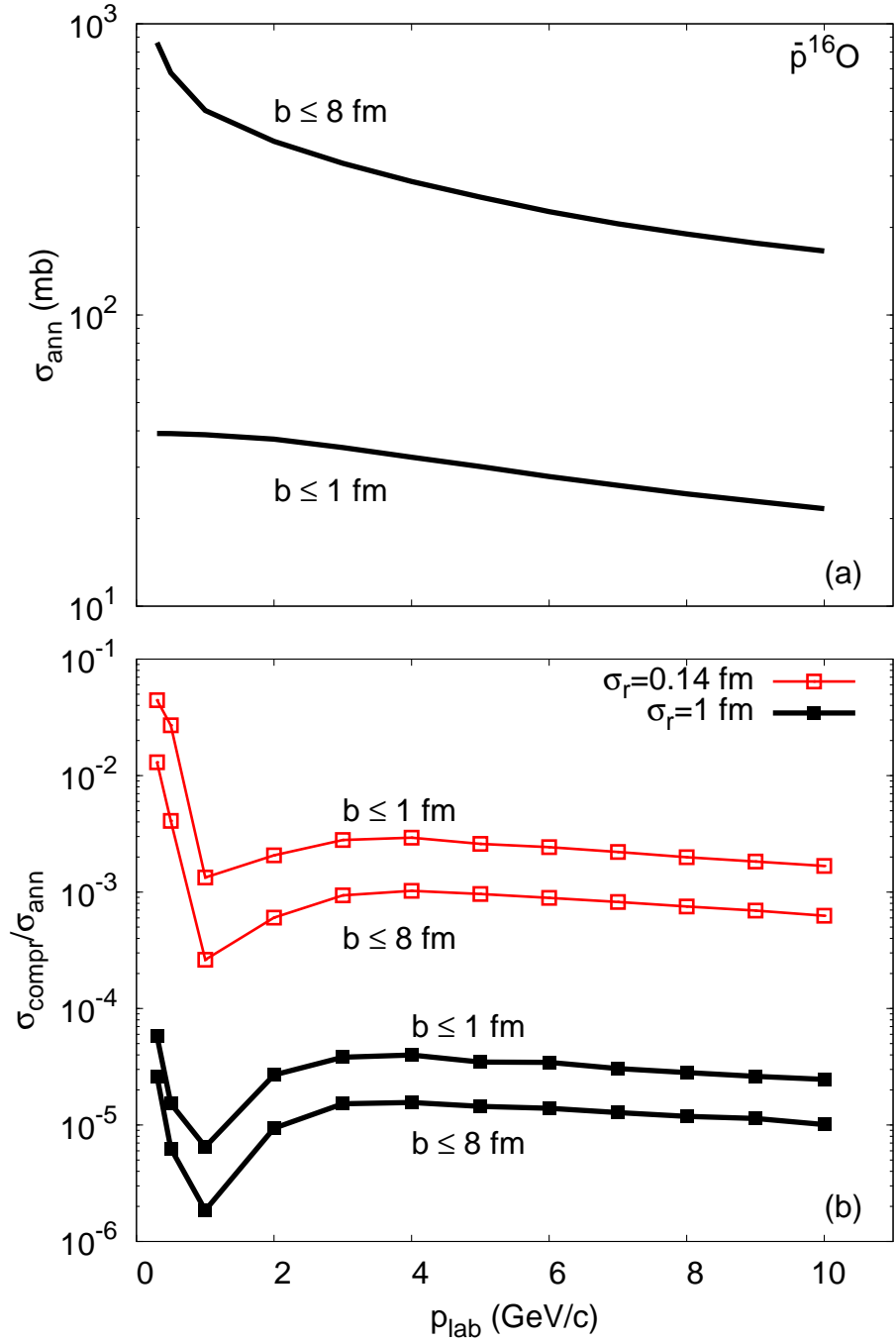}

\vspace*{0.5cm}

\caption{\label{fig:sigCompr}(Color online) Antiproton annihilation 
cross section on $^{16}$O (a) and the probability of annihilation
at $\rho_{\rm max} > 2\rho_0$ (b) vs a beam momentum.
Different lines correspond to different centralities:
$b \leq 8$ fm --- inclusive set of events, $b \leq 1$ fm --- central events.
The cases of a narrow ($\sigma_r=0.14$ fm) and a wide ($\sigma_r=1$ fm) 
initial antiproton space distribution are depicted in panel (b) by 
lines with open and filled squares, respectively.}
\end{figure}
\begin{figure}
\includegraphics[scale = 0.80]{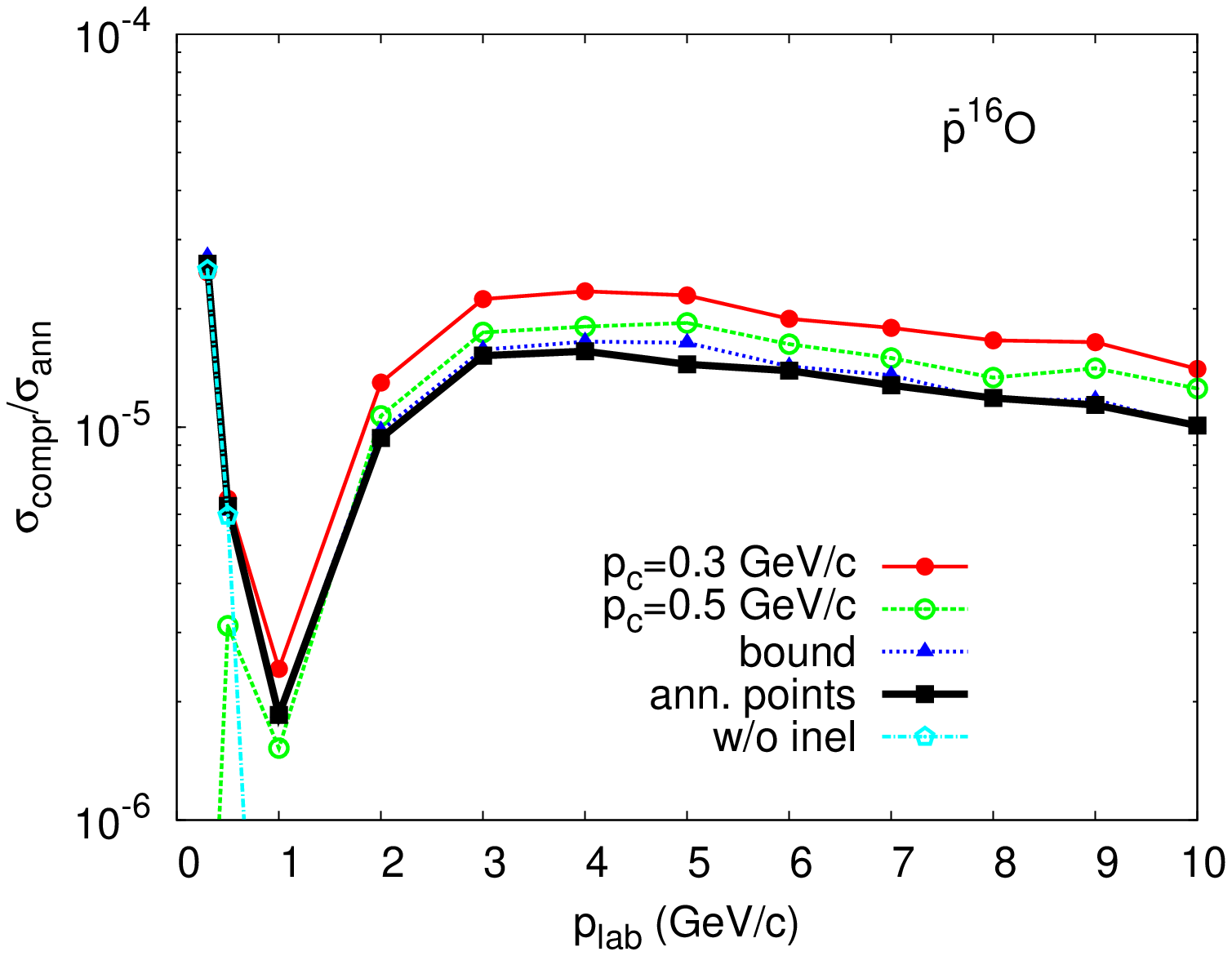} 

\vspace*{0.5cm}

\caption{\label{fig:sigCompr_cuts}(Color online)
Same as in Fig.~\ref{fig:sigCompr} (b) for the inclusive event set with $\sigma_r=1$ fm, 
but for different criteria of transition to the coherent compression dynamics (see text).
The lines with filled and open circles represent, respectively, 
the results obtained with a criterion using the critical momentum $p_c=0.3$ GeV/c and 
$0.5$ GeV/c, respectively. The line with filled triangles corresponds to the criterion
requiring that the antibaryon becomes bound. The line with filled squares shows the calculation 
with the default criterion using the annihilation points, same as in Fig.~\ref{fig:sigCompr}
(b). Additionally, the line with open pentagons shows the results obtained by 
switching-off the inelastic channels of the $\bar N N$ scattering.
In this case, the ACZ probability quickly drops with increasing beam momentum 
and becomes less than $10^{-8}$ at $p_{\rm lab} > 1$ GeV/c.}
\end{figure}

Figure~\ref{fig:sigCompr} shows the annihilation cross section $\sigma_{\rm ann}$ 
(a) and the relative fraction of ACZ $\sigma_{\rm compr}/\sigma_{\rm ann}$ (b)
as functions of the beam momentum.
While $\sigma_{\rm ann}$ drops with increasing $p_{\rm lab}$ due to the momentum
dependence of the $\bar p N$ annihilation cross section, the ratio 
$\sigma_{\rm compr}/\sigma_{\rm ann}$ reveals an interesting nonmonotonic
behaviour. First, it drops with increasing beam momentum up to 
$p_{\rm lab}\simeq1$ GeV/c and then starts to increase saturating at
$p_{\rm lab}\simeq3$ GeV/c. The growth of this ratio at $p_{\rm lab} > 1$ GeV/c 
is caused by opening the inelastic production channels,  
$\bar N N \to \bar N N \pi$ with the threshold beam momentum $p_{\rm thr}= 0.787$ GeV/c,
$\bar N N \to \bar N N \pi \pi$ with $p_{\rm thr}= 1.210$ GeV/c etc.
The inelastic production leads to the additional deceleration of an antibaryon and, 
therefore, increases the probability of the nuclear compression \cite{Mishustin:2004xa}
(see also Fig.~\ref{fig:sigCompr_cuts}).   
Selecting the central events increases the ratio $\sigma_{\rm compr}/\sigma_{\rm ann}$
by about a factor of three, which is caused by a larger relative fraction of 
annihilations at small radii (c.f. Fig.~\ref{fig:dsigdr}b).

The important result of the previous section is that only a slow
antiproton can induce nuclear compression. In practice, we have used
the ensemble of annihilation points to initialize the coherent GiBUU
runs assuming that antiprotons become slow not far away from their 
annihilation points. 
To check this assumption, we have performed additional calculations
with other transition criteria from collisional deceleration stage to the 
coherent compression dynamics. In the first calculation, we have generated
the ensemble of points where the momenta ${\bf p}$ and coordinates ${\bf r}$
of antibaryons satisfy the conditions $|{\bf p}| < p_c$ and $|{\bf r}| < r_c$
simultaneously, i.e. when the antibaryons become slow enough and close enough
to the nuclear centre. Here, $p_c$ and $r_c$ are parameters to be chosen. 
As follows from Fig.~\ref{fig:cont_psurv}, the choice $p_c\simeq0.3-0.5$ GeV/c,
$r_c\simeq3$ fm provides almost the full coverage of the $(r,p)$-region
where a significant ($\rho > 2\rho_0$) compression is expected. 
Another criterion selects the antibaryon momentum and position at the first 
time instant, when the antibaryon becomes bound, i.e. its energy falls below
its bare mass.
Figure~\ref{fig:sigCompr_cuts} shows the ACZ probability calculated by using the different
transition criteria. All results are quite similar, except for the calculation with 
$p_c=0.5$ GeV/c which becomes unphysical at $p_{\rm lab} < p_c$.

\begin{figure}[!t]
\includegraphics[scale = 0.80]{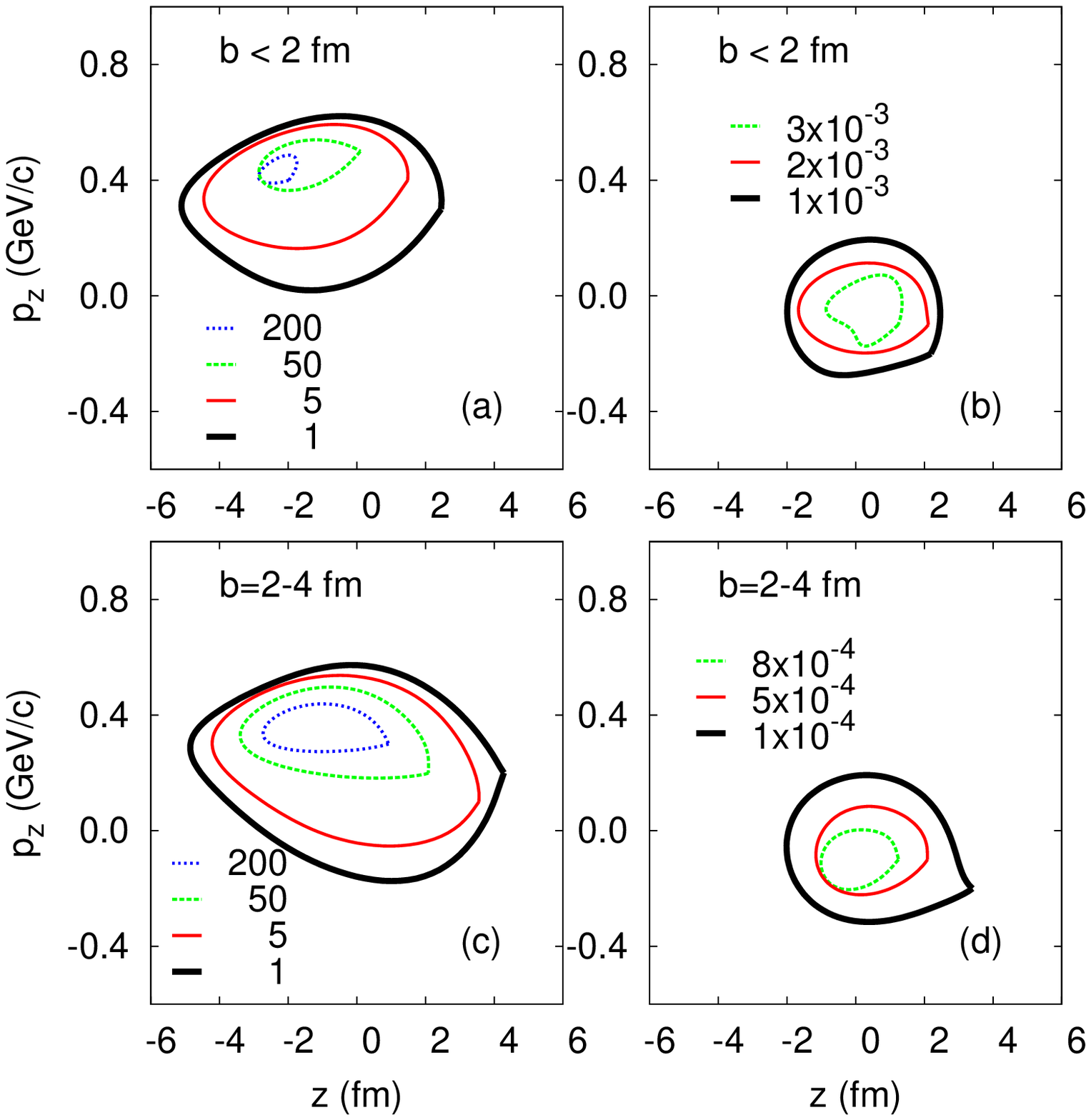}

\vspace*{0.5cm}

\caption{\label{fig:cont_z_pz_300mevc}(Color online) Panels (a) and (c) 
show the contour lines of the antiproton annihilation cross section 
$d\sigma_{\rm ann}/(dzdp_z)$ (mb/(fm $\cdot$ GeV/c)) in the plane 
of longitudinal coordinate $z$ and longitudinal momentum $p_z$ 
of the annihilation points for central (a) and peripheral (c) collisions.  
Panels (b) and (d) show the contour lines of relative fraction of 
annihilations at high density ($\rho_{\rm max} > 2\rho_0$), 
$\sigma_{\rm comp}/\sigma_{\rm ann}$, in the same plane for central
(b) and peripheral (d) collisions.
The colliding system is $\bar p ^{16}$O at 0.3 GeV/c.}
\end{figure}

\begin{figure}[!t]
\includegraphics[scale = 0.80]{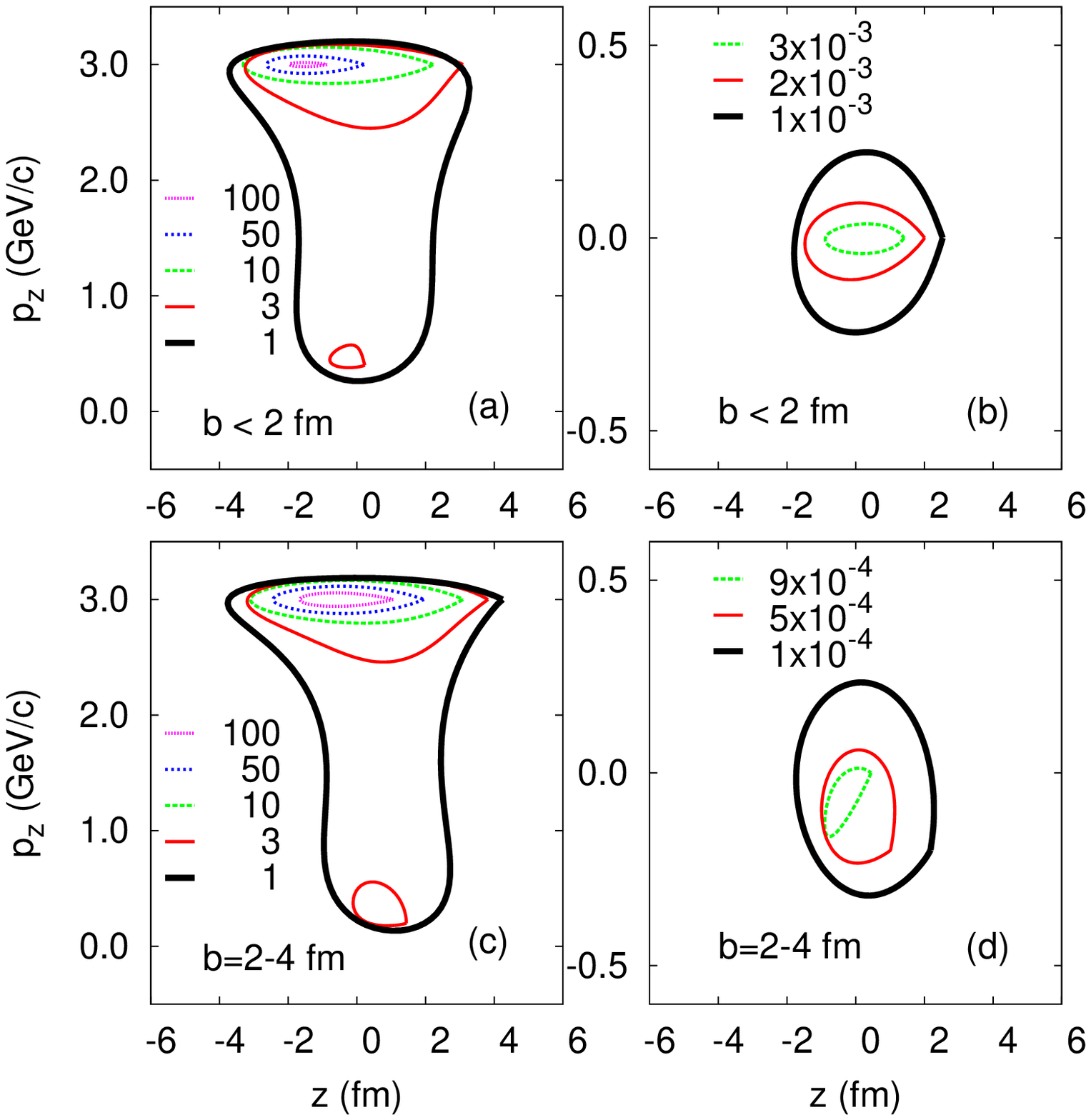}

\vspace*{0.5cm}

\caption{\label{fig:cont_z_pz_3000mevc}(Color online)
Same as Fig.~\ref{fig:cont_z_pz_300mevc}, but for $p_{\rm lab}=3$ GeV/c.
Notice different scales of vertical axes in the left and
right panels.}
\end{figure}

To give more insight into the $\bar p$-induced nuclear compression, in
Figs.~\ref{fig:cont_z_pz_300mevc} and \ref{fig:cont_z_pz_3000mevc} we show the 
distributions of annihilation events on the longitudinal coordinate $z$ and 
the longitudinal momentum $p_z$ of the antibaryon for central (a)
and peripheral (c) $\bar p ^{16}$O collisions. The relative probability
of annihilation in the compressed zone as a function of $z$ and $p_z$ is
shown in panels (b) and (d) for central and peripheral collisions, 
respectively. Independent of the beam momentum, the maximum ACZ probability
is reached if the antibaryon is stopped in the central region. 
However, the longitudinal coordinates for events most favourable 
for compression are rather uncertain, as expected already from Fig.~\ref{fig:rhoz}.
On the other hand, we definitely observe a rather strong impact parameter dependence
with the clear preference of central collisions for selecting the ACZ events.
At large beam momenta (Fig.~\ref{fig:cont_z_pz_3000mevc}), the compression can only 
be reached at the extreme tail of the antibaryon longitudinal momentum 
distribution, and the total probability of ACZ is small. As one can see from the 
right panels in Figs.~\ref{fig:cont_z_pz_300mevc} and \ref{fig:cont_z_pz_3000mevc},
a significant compression ($\rho_{\rm max} > 2\rho_0$) can be produced by antibaryons
whose longitudinal momenta are less than 200 MeV/c. However, the maximum relative 
probability of ACZ in the $(z,p_z)$-plane is practically independent on the beam momentum.
This is expected, since in our model the probability of ACZ depends only on the 
position and momentum of the antibaryon prior the annihilation.

Since nuclear compression is most probable for stopped annihilations, 
one needs a trigger to select the events with slow antiprotons.
We will discuss two possible triggers here.

The first trigger requires the detection of a fast proton in forward direction 
\cite{PANDA}. The idea behind is that the incoming antiproton can be
decelerated and captured in a nucleus by experiencing a hard collision with a 
single nucleon.
\begin{figure}
\includegraphics[scale = 0.80]{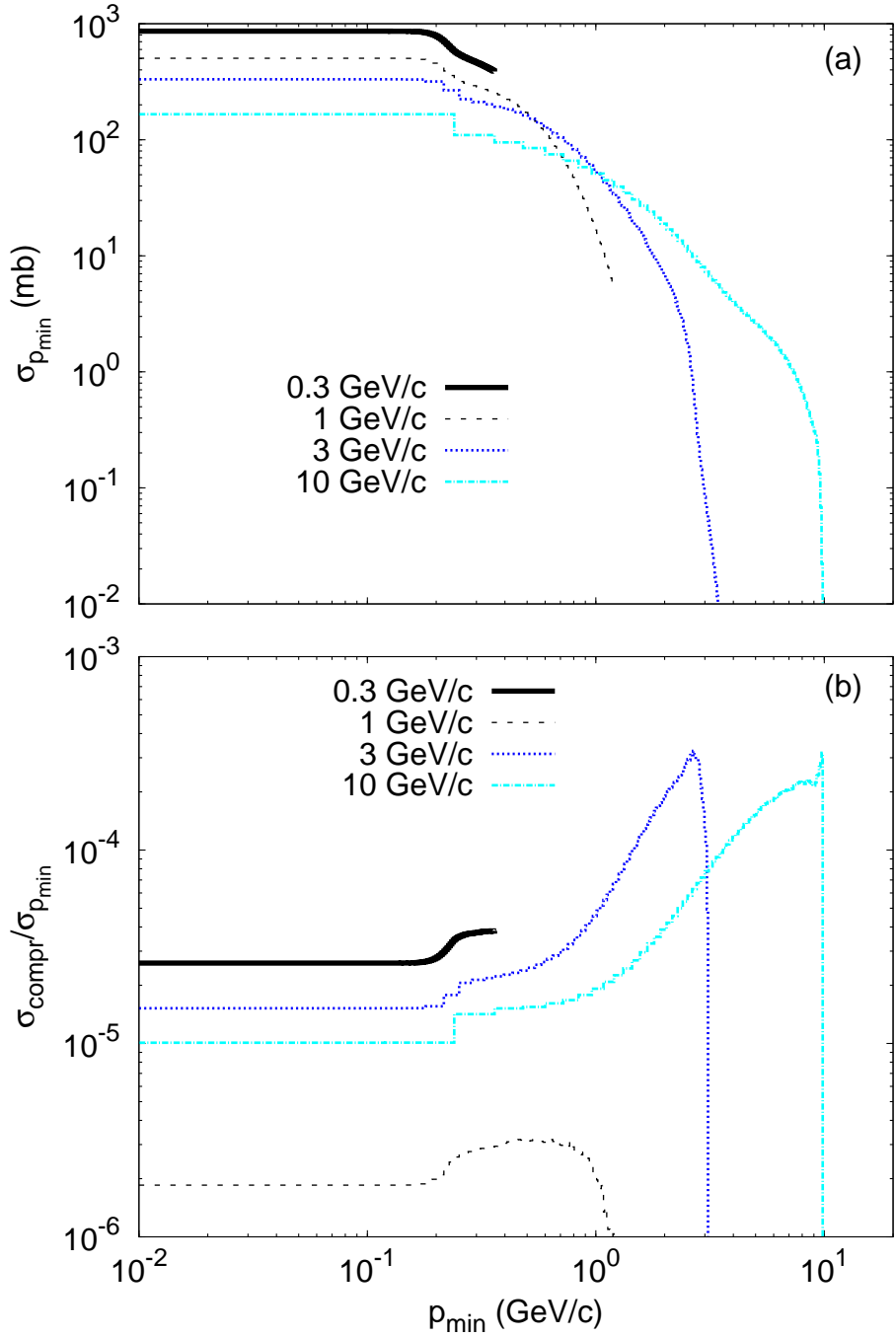}

\vspace*{0.5cm}

\caption{\label{fig:sigCompr_pfast}(Color online) (a) The cross section $\sigma_{p_{\rm min}}$ 
of $\bar p$-annihilation on $^{16}$O accompanied by the emission of a proton
with momentum larger than $p_{\rm min}$ as a function of $p_{\rm min}$.
(b) The relative probability $\sigma_{\rm compr}/\sigma_{p_{\rm min}}$ of 
the annihilation in compressed zone  $(\rho_{\rm max} > 2\rho_0)$ vs the minimum
momentum $p_{\rm min}$ of the emitted proton. Different curves refer to different
$\bar p$-beam momenta.}  
\end{figure}
Figure~\ref{fig:sigCompr_pfast} (a) shows the cross section $\sigma_{p_{\rm min}}$ 
of an antiproton annihilation on $^{16}$O accompanied by the emission of a proton
with momentum exceeding some value $p_{\rm min}$ as a function of $p_{\rm min}$.
For simplicity, we did not apply any angular cuts for the proton momentum.
At large beam momenta, 3 and 10 GeV/c, the cross section $\sigma_{p_{\rm min}}$
sharply drops with $p_{\rm min}$ near $p_{\rm min} \simeq p_{\rm lab}$.
In the panel (b) of Fig.~\ref{fig:sigCompr_pfast}, we show the relative fraction 
$\sigma_{\rm compr}/\sigma_{p_{\rm min}}$ of ACZ as a function of the minimum proton 
momentum $p_{\rm min}$.
For $p_{\rm lab}=3$ and 10 GeV/c, the quantity $\sigma_{\rm compr}/\sigma_{p_{\rm min}}$
grows by almost a factor of thirty while $p_{\rm min}$ increases from zero to $p_{\rm lab}$.  

Emission of a fast proton with momentum close to the $\bar p$-beam momentum can be caused by the 
following mechanisms: (i) elastic scattering $\bar p + p \to \bar p_{\rm slow} + p_{\rm fast}$, 
(ii) inelastic production processes of the type
$\bar p + p \to \bar p_{\rm slow} + p_{\rm fast}+{\rm mesons}$,
and (iii) collisions with high-momentum annihilational pions, $\pi+ p \to p_{\rm fast} + X$.
We have checked, that inelastic reactions (ii) give the largest contribution to the production 
of the fast proton at $p_{\rm lab}=3$ and 10 GeV/c. This makes the fast proton trigger rather
efficient at high beam momenta. 
The contribution from process (iii) is relatively small, while elastic scattering (i)
practically does not contribute to the yield of fast protons at $p_{\rm lab}=3$ and 10 GeV/c.   
On the other hand, for small beam momenta, 0.3 and 1 GeV/c, the pionic mechanism (iii)
contributes dominantly to the fast proton yield with only a small admixture of elastic
scattering (i). Therefore, the trigger based on a fast proton in final state is 
ineffective at small beam momenta.

\begin{figure}
\includegraphics[scale = 0.80]{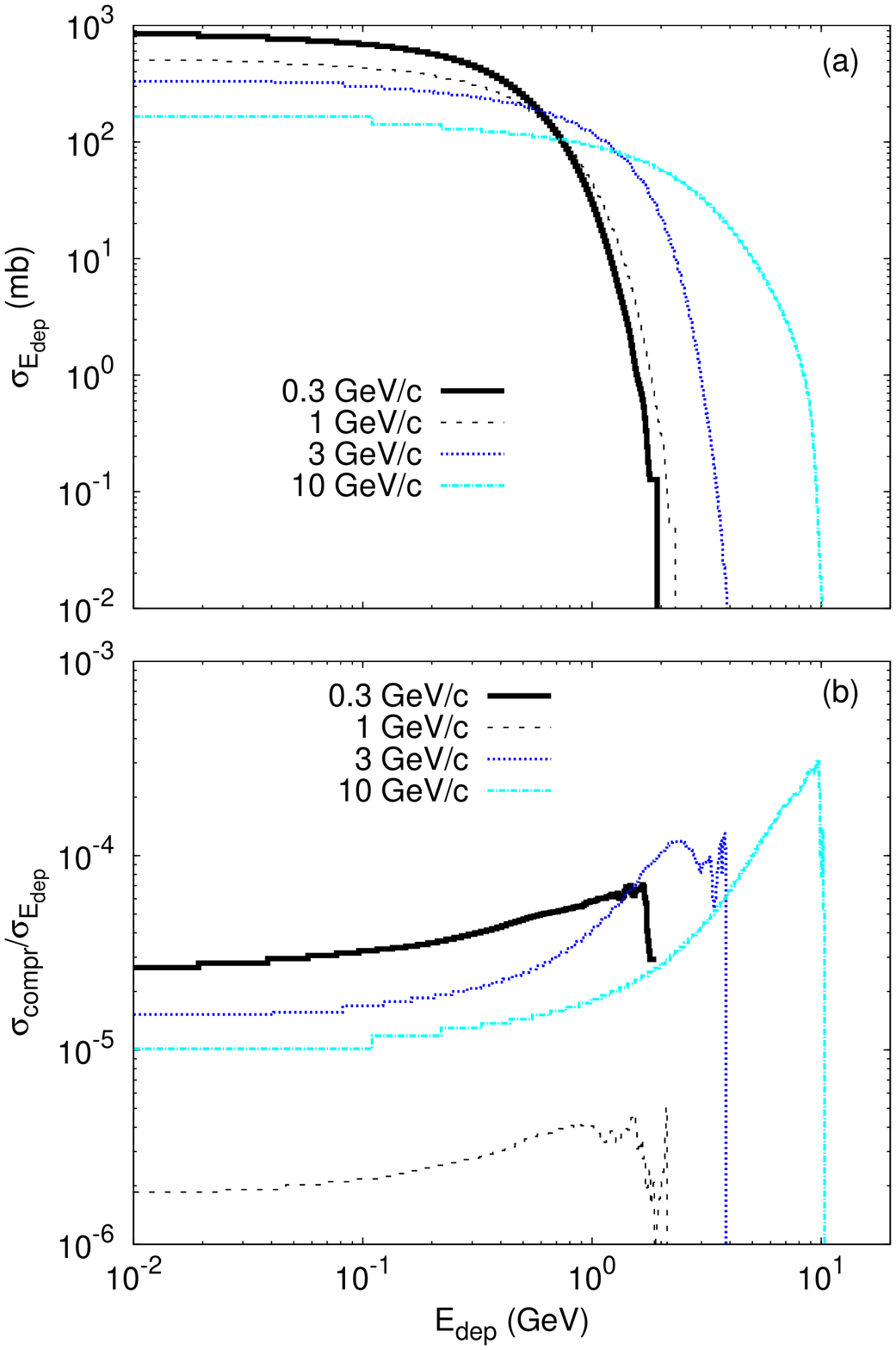}

\vspace*{0.5cm}

\caption{\label{fig:sigCompr_edep}(Color online) (a) Antiproton annihilation 
cross section on $^{16}$O at the condition of the energy deposition 
exceeding $E_{\rm dep}$ as a function of $E_{\rm dep}$.
(b) The relative probability $\sigma_{\rm compr}/\sigma_{E_{\rm dep}}$
of a compressional annihilation $(\rho_{\rm max} > 2\rho_0)$ as 
a function of $E_{\rm dep}$. Different curves correspond to different
$\bar p$-beam momenta.}  
\end{figure}

The second possible trigger is based on the energy deposition
\cite{Clover:1982qq,Cugnon:1986tx}
\begin{equation}
   E_{\rm dep} = T_{\bar p} + 2m_N - \sum_{i} E_{\rm mes}^{(i)}~, \label{E_dep}
\end{equation}
where $T_{\bar p}$ is the antiproton beam energy, $E_{\rm mes}^{(i)}$ is the energy
of $i$-th outgoing meson, and the sum runs over all produced mesons in 
a given event. Neglecting nucleon and antibaryon binding energy, 
antiproton elastic and inelastic scattering before annihilation and final state 
interactions of produced mesons, one has $E_{\rm dep}=0$. In the case of low energy 
antiproton-nucleus collisions, annihilations with a larger energy deposition occur 
deeper in the nucleus, as was found in \cite{Clover:1982qq}. The explanation was
that the annihilation mesons loose their energy or get absorbed more effectively
if the annihilation takes place deeply inside the nucleus.
For high-energy $\bar p$-nucleus interactions, the incoming antiproton can rescatter 
before annihilation transferring a part of energy to the nucleons. This also leads to 
larger values of $E_{\rm dep}$, since the produced mesons will have a smaller total 
energy in this case. 
Both types of events, deep and/or slow antibaryon annihilations,
should be of the ACZ-type with an increased probability. Results for $E_{\rm dep}$-trigger
are shown in Fig.~\ref{fig:sigCompr_edep}. As one can see, triggering on a large energy
deposition, $E_{\rm dep} \simeq T_{\bar p} + 2m_N$, increases  the fraction of ACZ events
by about a factor of thirty with respect to $E_{\rm dep}\simeq0$ for the beam momentum 
of 10 GeV/c.   

\section{In--medium modifications of antiproton annihilation}
\label{Inmedium}

So far we have used the vacuum $\bar p$N annihilation cross section and the fixed antibaryon
mean field parameters. As the survival probability of an antiproton (\ref{Psurv}) is determined
by its annihilation width (\ref{Gamma_ann}), it is important to consider possible
in-medium modifications of the latter. On the other hand, the speed and the amplitude of the 
nuclear compression depends on the value of the reduction factor $\xi$ of the antibaryon-meson
coupling constants \cite{Larionov:2008wy}. Thus, the probability of ACZ is the result of the
competition between compression and annihilation. In this section, we discuss possible 
modifications of the antiproton annihilation width in nuclear medium.
The sensitivity of our results to the in-medium modifications is demonsrated in 
Fig.~\ref{fig:probCompr_vs_rho} below.

As discussed by many authors (see e.g. Refs.~\cite{Mishustin:2004xa,Rafelski:1979nt,
Cugnon:1984zp,Cugnon:1989hj,Kahana:1992hx,Pang:1996ez,Spieles:1995fs,Hernandez:1986ev,
Hernandez:1989ij,Hernandez:1992rv}), the annihilation rate of antiprotons in a dense 
nuclear medium may significantly differ from simple calculations using the vacuum 
$\bar p$N annihilation cross section $\sigma_{\rm ann}$ (see Eq.~(\ref{Gamma_ann})).
There are several effects which become important at sufficiently high nucleon densities. 
First, the dispersion relations of nucleons and antinucleons are modified 
due to interactions with mean mesonic fields.  
In particular, the effective masses $m_N^*$ and $m_{\bar N}^*$ are reduced compared 
to the vacuum value. 
Generally, this leads to the reduction of the imaginary parts of the 
nucleon and antinucleon self-energies in nuclear medium.
The influence of the baryon and antibaryon in-medium dispersion relations on the antibaryon 
annihilation rate has been studied in Ref.~\cite{Mishustin:2004xa}.
Other examples of the influence of the baryonic effective masses on 
hadronic processes are the in-medium reduced cross sections of the 
NN elastic scattering \cite{Pandharipande:1992zz,Fuchs:2001fp} 
and of the $\Delta$-resonance production $NN \to N\Delta$ \cite{TerHaar:1987ce,
Larionov:2003av}. As an illustrative example of the in-medium reduction 
caused by effective masses, we consider the two-pion annihilation channel 
later-on in this section.
  
Another important in-medium effect is the appearance of the MNA channels. 
The famous Pontecorvo reaction \cite{Pontecorvo:1956vx} $\bar p d \to \pi^- p$
is an example of the MNA processes. 
It is commonly believed that MNA is responsible for the emission of 
high-energy protons from low-energetic $\bar p$ annihilation on nuclei 
\cite{Hernandez:1989ij,Minor:1992,Montagna:2002wg}. Moreover, the triggering
on high-momentum protons is already applied in experimental techniques to distinguish 
MNA from the single-nucleon annihilation (SNA) followed by the final state 
interaction (rescattering and absorption) of produced mesons 
\cite{Montagna:2002wg}. 

Up to now the attempts to estimate the MNA contribution performed by different theoretical 
and experimental groups did not lead to definite conclusions. 
The experimental determinations of the MNA probability give the values  
of about 10-30\% for the $\bar p$ annihilations at rest \cite{Minor:1992,Montagna:2002wg}.
One has to admit that these values agree with predictions of Hern\'andez and Oset (HO) 
\cite{Hernandez:1986ev,Hernandez:1989ij,Hernandez:1992rv}.
It is important for this agreement, however, that the annihilations of stopped antiprotons 
take place {\it at the nuclear fringe}, $\rho \sim 0.1 \rho_0$ \cite{Minor:1992}.
On the other hand, HO argue in Ref.~\cite{Hernandez:1989ij} that the MNA
channels are required to describe the high-momentum tails of the proton momentum 
spectra from $\bar p$ annihilation at $p_{\rm lab}=608$ MeV/c on $^{12}$C \cite{Mcgaughey:1986kz}.
However, the intranuclear cascade calculations \cite{Cugnon:1989mx}
and the GiBUU calculations \cite{Larionov:2009tc} have demonstrated that
the agreement with the data \cite{Mcgaughey:1986kz} can be reached without 
any unusual mechanisms.

Using a diagram language, HO have considered $\bar p$N annihilation vertices
including virtual pions which may decay into particle-hole excitations  
\cite{Hernandez:1986ev,Hernandez:1989ij,Hernandez:1992rv}.
These diagrams can be interpreted as MNA channels, which, according to 
the HO calculations, have extremely high probability at $\rho \sim \rho_0$,
one order of magnitude higher than the ordinary SNA channels. 
This result is in a clear contradiction with the theoretical estimates by 
Cugnon and Vandermeulen \cite{Cugnon:1984zp,Cugnon:1989hj} and by Mishustin 
et al. \cite{Mishustin:2004xa}, although these estimates are based 
on relatively simple statistical considerations.
In our opinion, the HO calculations have a problem with convergence of the series 
in powers of $\rho$ at $\rho \geq \rho_0$ (Eq. (4.3) in Ref.~\cite{Hernandez:1989ij}). 
Since it is very difficult to test the MNA probability at $\rho \sim \rho_0$ experimentally, 
different theoretical predictions are still possible here.

%Here, we would like to emphasize that the mechanism considered
%in Refs.~\cite{Hernandez:1986ev,Hernandez:1989ij,Hernandez:1992rv} is to a large
%extent equivalent to a simple geometrical picture of MNA proposed by other
%authors. 
%Indeed, cutting intermediate lines in the antiproton self-energy diagram 
%(2b) in Ref.~\cite{Hernandez:1989ij} makes it similar to
%the (1b)-type diagram introduced earlier in the same paper. 
%But the latter diagram is nothing but the annihilation channel involving
%an additional nucleon. Its contribution can be estimated directly using simple 
%geometric considerations of Ref.~\cite{Mishustin:2004xa}.

In Ref.~\cite{Mishustin:2004xa}, the relative importance of MNA 
channels was evaluated by calculating the probability to find more than 
one nucleon in the annihilation volume $V_{\rm ann}$. 
This calculation was done for a spherical volume with the radius $R_{\rm ann} \simeq 0.8$ fm 
assuming the Poisson distribution in the number of nucleons $n$,
$P(n)=\overline{n}^{\,n} \exp(-\overline{n})/n!$, where $\overline{n}=\rho V_{\rm ann}$ 
is the average number of nucleons in this volume. 
In the case of enhanced density, $\rho \simeq 2\rho_0$, one has $\overline{n}\simeq 0.6$.
This leads to the probability of MNA channels with $n>1$ on the level of 40\% 
of the SNA ($n=1$), which is about one order of magnitude smaller than the value
predicted in Refs.~\cite{Hernandez:1986ev,Hernandez:1989ij,Hernandez:1992rv}.

In the literature, one can also find other arguments against a large enhancement 
of $\bar p$N annihilation cross section in nuclear medium. For instance, as argued in 
Refs.~\cite{Kahana:1992hx,Pang:1996ez}, the presence of additional nucleon(s) in the 
annihilation volume may lead to the ''screening'' of in-medium annihilation as compared 
to the usual SNA mechanism. By introducing the screening effect these authors achieve 
a better agreement with experimental data on $\bar p$ production in $pA$ and 
$AA$ collisions at AGS energies. 

In order to illustrate the influence of the in-medium effective masses on the antiproton 
annihilation, let us consider a relatively simple case of the two-pion annihilation 
$\bar p p \to \pi^- \pi^+$.
In vacuum, this process can be described by the one-nucleon exchange model 
\cite{Moussallam:1984uj,Moussallam:1984zj}. In the Born approximation,
the matrix element can be written as follows:
\begin{equation}
   M = -2 F^2(t) 
\bar v({m_s}_{\bar p},p_{\bar p}) [A - \not{k} B(t)] u({m_s}_p,p_p) ~,   \label{matrElem}
\end{equation}
where $u$ and $v$ are, respectively, the proton and antiproton bispinors
($\bar u u = 1$, $\bar v v = -1$), which depend on the spin magnetic quantum numbers
${m_s}_p,~{m_s}_{\bar p}$ and on the four-momenta $p_p,~p_{\bar p}$, and
$k$ is the four-momentum of $\pi^+$.
The scalar parameters $A$ and $B$ are defined as
\begin{equation} 
   A=\frac{f^2}{m_\pi^2}2m_N,
~~~B(t)=\frac{f^2}{m_\pi^2}\left(1+\frac{4m_N^2}{t-m_N^2}\right)~,  \label{AB}
\end{equation}
where $f=1.008$ is the pion-nucleon coupling constant and $t=(p_p-k)^2$. 
The factor of -2 in (\ref{matrElem}) is obtained from an isospin algebra. 
The off-shell nucleon form factor is chosen as in \cite{Moussallam:1984uj}:
\begin{equation}
   F(t)=\left(\frac{\Lambda^2-m_N^2}{\Lambda^2-t}\right)^{1/2}~,    \label{formFactor}
\end{equation}
where $\Lambda$ is a cut-off parameter. In the center-of-mass (c.m.) frame, 
the differential cross section of the process $\bar p p \to \pi^- \pi^+$ is
\begin{equation}
   \frac{d \sigma_{\bar p p \to \pi^- \pi^+}}{d \Omega_{c.m.}}
   = \frac{(2m_N)^2}{64 \pi^2 s}
     \overline{|M|^2} 
     \frac{q_{\pi\pi}}{q_{\bar p p}}~,    \label{dsigdOmega}
\end{equation}
where $s=(p_{\bar p}+p_p)^2$, $q_{\bar p p}=q(\sqrt{s},m_N)$ and
$q_{\pi\pi}=q(\sqrt{s},m_\pi)$ are the c.m. momenta of the incoming
and outgoing particles, respectively, with $q(\sqrt{s},m)=(s/4-m^2)^{1/2}$,
and $\overline{|M|^2}=\frac{1}{4} \sum\limits_{{m_s}_{\bar p},{m_s}_p} |M|^2$.
\begin{figure}
\includegraphics[scale = 0.80]{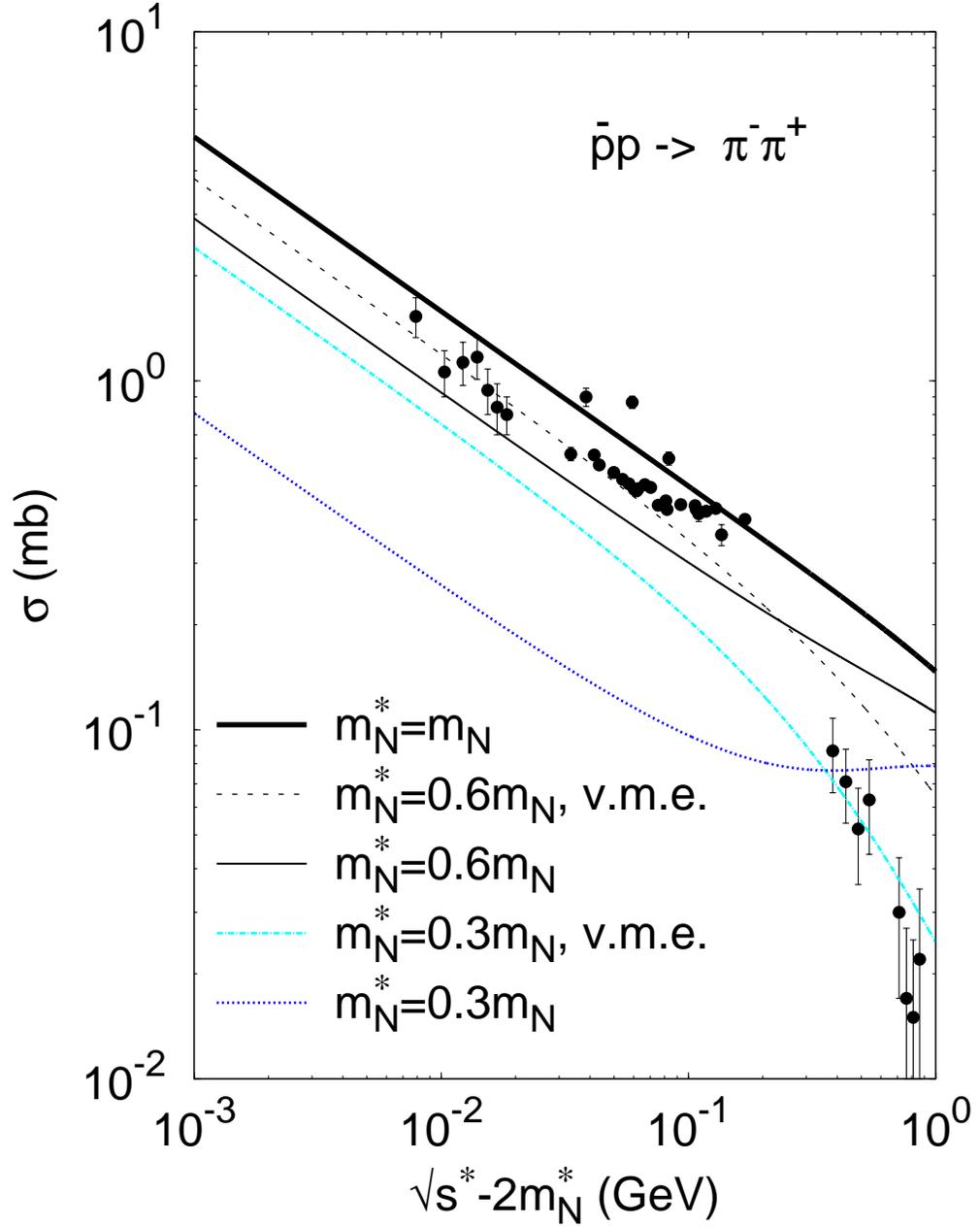}

\vspace*{0.5cm}

\caption{\label{fig:ppbar_pipi} The cross section of
the process $\bar p p \to \pi^- \pi^+$ as a function of the
total in-medium c.m. kinetic energy $\sqrt{s^*}-2m_N^*$.
The results are shown for the different choices of a nucleon
effective mass $m_N^*$. The calculation with $m_N^*=m_N$ (thick solid line) 
corresponds to the vacuum cross section, which has to be compared with 
experimental data. The calculations with $m_N^*=0.6m_N$ 
(thin solid line --- full result,
dashed line --- vacuum matrix element)  
and $m_N^*=0.3m_N$ (dotted line --- full result,
dash-dotted line --- vacuum matrix element) 
represent the in-medium cross sections. 
Experimental data are from Refs. \cite{Eastman:1973va,
Tanimori:1985ps,Bardin:1987hw,Sugimoto:1987ct,Bardin:1994am}.}  
\end{figure}
Solid line in Fig.~\ref{fig:ppbar_pipi} reports the total vacuum 
$\bar p p \to \pi^- \pi^+$ cross section calculated in the Born approximation.
This is a quite rough approximation in the case of $\bar p p$ incoming 
channel. We stress, however, that our main purpose here is just to 
demonstrate the influence of the in-medium effects and not to perform 
the state-of-art calculations for the two-pion annihilation in vacuum. 
To fit the data for slow antiprotons ($p_{\rm lab} < 1$ GeV/c), we have 
chosen a rather small value of the cut-off parameter $\Lambda=1.0$ GeV,
since the initial state interactions are neglected 
(see discussion in Refs. \cite{Moussallam:1984uj,Moussallam:1984zj}). 

Assuming for simplicity the G-parity transformed proton scalar and vector potentials
acting on the antiproton, the baryonic mean fields can be now taken into account by 
replacing $m_N \to m_N^*$, $p_p \to p_p^*$, $p_{\bar p} \to p_{\bar p}^*$, $s \to s^*=(p_p^* + p_{\bar p}^*)^2$,
and $t \to t^*=(p_p^*-k)^2$ in Eqs. (\ref{matrElem}),(\ref{AB}) and (\ref{dsigdOmega})
(c.f. Refs. \cite{Kim:1996ada,Larionov:2003av}). Note, that we always keep 
the vacuum nucleon mass in the numerator $\Lambda^2-m_N^2$ of the form factor
(\ref{formFactor}), since the above replacements are motivated by the baryon 
in-medium self-energies which should not change the form factor fixed in vacuum.
Then, the total in-medium $\bar p p \to \pi^- \pi^+$ cross section reads: 
\begin{equation}
   \sigma_{\bar p p \to \pi^- \pi^+}^{\rm med}(\sqrt{s^*})
  =\frac{(2m_N^*)^2 q(\sqrt{s^*},m_\pi)}{32 \pi s^* q(\sqrt{s^*},m_N^*)}
   \int\limits_{-1}^1\,d\cos\,\Theta_{c.m.}\,
   \overline{|M^{\rm med}|^2}(\sqrt{s^*},\cos\,\Theta_{c.m.})~.
                                                   \label{sig_pbarp2pi-pi+^med}
\end{equation}
Here, the quantity $M^{\rm med}$ is the in-medium matrix element, while
$\Theta_{c.m.}$ is the angle between the proton and $\pi^+$ three-momenta
in the c.m. frame. The results of calculation using Eq. (\ref{sig_pbarp2pi-pi+^med})
at $m_N^*=0.6m_N$ ($\rho=\rho_0$) and $m_N^*=0.3m_N$  ($\rho\simeq2\rho_0$) are shown in 
Fig.~\ref{fig:ppbar_pipi} by the thin solid and dotted lines, respectively. 
As one can see, the in-medium $\bar p p \to \pi^- \pi^+$ cross section is strongly reduced,
largely due to the $(2m_N^*)^2$ factor in Eq. (\ref{sig_pbarp2pi-pi+^med}), which 
comes from the in-medium Dirac bispinor normalization.
\begin{figure}
\includegraphics[scale = 0.80]{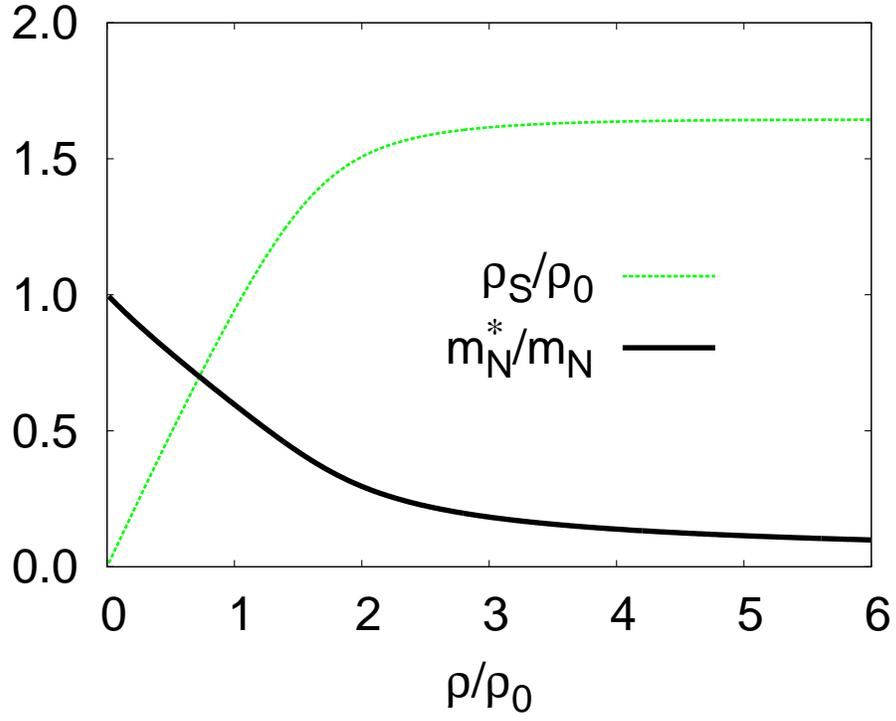}

\vspace*{0.5cm}

\caption{\label{fig:mstar_vs_rho}(Color online) Nucleon effective mass and
scalar density vs the baryon density in infinite nuclear matter in the case of 
NL3 interaction \cite{Lalazissis:1996rd} applied in the present work.}  
\end{figure}
For orientation, we present in Fig.~\ref{fig:mstar_vs_rho}
the baryon density dependence of the nucleon effective mass $m_N^*$ and of the nucleon scalar
density $\rho_S$ (see Eqs. (8),(9) and (15) in Ref. \cite{Larionov:2008wy}).
The effective mass $m_N^*$ drops with increasing baryon density which is an
important effect influencing in-medium cross sections (see also 
Refs. \cite{Pandharipande:1992zz,TerHaar:1987ce,Fuchs:2001fp,Larionov:2003av}.)

The two-pion annihilation channels represent, however, less than 1\% of the total
annihilation cross section, while the direct calculation of multi-meson annihilation
matrix elements is impossible. We will assume that the matrix elements are not
modified in nuclear medium and take into account only the in-medium bispinor
normalization, flux and phase space volume. This intuitive assumption
has some support from the earlier studies of $NN \to NN$ and $NN \to N\Delta$
cross sections in nuclear matter 
\cite{Pandharipande:1992zz,TerHaar:1987ce,Fuchs:2001fp,Larionov:2003av}.
In this way, one can write the following formula for the in-medium
cross section of the $\bar N N \to M_1,M_2,...,M_n$ annihilation channel with
$n$ outgoing mesons (c.f. \cite{Wagner:2004ee,Larionov:2007hy,Larionov:2008wy}):
\begin{eqnarray}
   \sigma_{\bar N N \to M_1,M_2,...,M_n}^{\rm med}(\sqrt{s^*})
    &=& \sigma_{\bar N N \to M_1,M_2,...,M_n}(\sqrt{s_{\rm corr}})
  \left(\frac{m_N^*}{m_N}\right)^2
  \frac{I_{\bar N N}}{I_{\bar N N}^*}  \nonumber \\ 
 &\times& \frac{\Phi_n(\sqrt{s^*};m_1,m_2,...,m_n)}%
  {\Phi_n(\sqrt{s_{\rm corr}};m_1,m_2,...,m_n)}~. \label{sigMed}
\end{eqnarray}
Here $m_1,m_2,...,m_n$ are the vacuum masses of outgoing mesons, 
$\sqrt{s_{\rm corr}}=\sqrt{s^*}-2(m_N^*-m_N)$ is the so-called corrected
invariant energy of the colliding particles, the analogue of the invariant
energy in vacuum. The quantities $I_{\bar N N}=q(\sqrt{s_{\rm corr}},m_N)\sqrt{s_{\rm corr}}$ and 
$I_{\bar N N}^*=q(\sqrt{s^*},m_N^*)\sqrt{s^*}$ are the vacuum and in-medium flux
factors, respectively. The $n$-body phase space volume is defined as
\begin{eqnarray}
   \Phi_n(\sqrt{s};m_1,m_2,...,m_n)
    & = & \int\,\frac{ d^3 k_1 }{ (2\pi)^3 2\omega_1 } 
      \int\,\frac{ d^3 k_2 }{ (2\pi)^3 2\omega_2 } \cdots 
      \int\,\frac{ d^3 k_n }{ (2\pi)^3 2\omega_n }     \nonumber \\  
    & \times &  \delta^{(4)}({\cal P} - k_1 - k_2 - ... - k_n)~,        \label{Phi_n}
\end{eqnarray}
where ${\cal P}^2=s$ and $\omega_i^2-{\bf k}_i^2=m_i^2$ ($i=1,2,...,n$).
The quantity $\sigma_{\bar N N \to M_1,M_2,...,M_n}(\sqrt{s_{\rm corr}})$ is
the vacuum cross section. Application of Eq. (\ref{sigMed}) to the process
$\bar p p \to \pi^- \pi^+$ leads to the formula (\ref{sig_pbarp2pi-pi+^med})
with replacement $\overline{|M^{\rm med}|^2}(\sqrt{s^*},\cos\,\Theta_{c.m.})
\to \overline{|M|^2}(\sqrt{s_{\rm corr}},\cos\,\Theta_{c.m.})$. The results obtained
using Eq. (\ref{sigMed}) for $m_N^*=0.6m_N$ and $m_N^*=0.3m_N$ 
are shown in Fig.~\ref{fig:ppbar_pipi} by the dashed 
and dash-dotted lines, respectively. The conclusion is that 
using vacuum matrix element produces somewhat less
pronounced in-medium reduction of the cross section.

In order to simulate the mean field effects on the total annihilation cross section,
we represent it as a sum of partial annihilation cross sections with various outgoing 
mesonic channels. In practice, this is done by using the statistical annihilation model 
of Refs. \cite{Golubeva:1992tr,PshenichnovPhD}.
Then we apply Eq. (\ref{sigMed}) to every annihilation channel with up to $n=6$
outgoing mesons. The mean field effects on the annihilation channels with more than six 
outgoing mesons are neglected.

In the following GiBUU calculations of the present section, we keep the first stage 
(see Sect.~\ref{Procedure}) unchanged: it is always computed with the reduction factor $\xi=0.22$ and 
vacuum $\bar p$N annihilation cross sections.
This is reasonable, since the compressional response of a nuclear system 
on a fast moving antiproton is weak and can be neglected.
On the other hand, the in-medium corrections to the $\bar p$N annihilation cross
section should also weaken for the fast antiproton. Thus, we vary the model
parameters for the second stage compressional dynamics only.

\begin{figure}
\includegraphics[scale = 0.80]{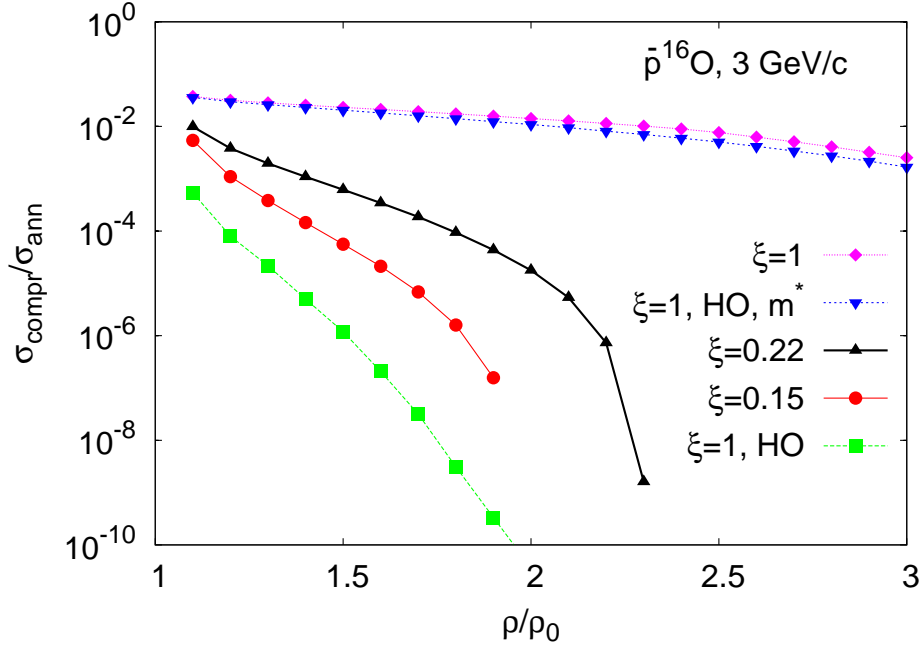}

\vspace*{0.5cm}

\caption{\label{fig:probCompr_vs_rho}(Color online) Probability of $\bar p$ annihilation
at $\rho_{\rm max} > \rho$ vs $\rho$ for $\bar p$$^{16}$O collisions
at $p_{\rm lab}=3$ GeV/c. The calculations with various values of the reduction
factor $\xi$ and vacuum $\bar p$N annihilation  cross section
are shown by lines with points denoted by the values of $\xi$ only.
The calculation with the in-medium enhanced $\bar p$N annihilation probability 
by the HO formula (Eq. (5) in Ref. \cite{Hernandez:1992rv} with parameters
taken at the $\bar p$ kinetic energy of 50 MeV) and $\xi=1$ are 
shown by the dashed line with filled boxes.
The result taking into account both the HO formula and the mean field reduction
of the $\bar p$N annihilation cross section according to Eq. (\ref{sigMed}) 
is shown by the dashed line with upside-down filled triangles.
The calculations for $\rho \leq \rho_0$ are not shown, since they are influenced
by the finite size of the $(r,p,\cos\Theta)$ grid in the space of the antiproton 
radial position $r$, momentum $p$ and $\cos\Theta={\bf r p}/rp$ 
(see Sect.~\ref{Procedure}).}  
\end{figure}
Fig.~\ref{fig:probCompr_vs_rho} shows the probability of $\bar p$ annihilation
at $\rho_{\rm max} > \rho$ as a function of $\rho$ for the inclusive set of
$\bar p$$^{16}$O events at the beam momentum of 3 GeV/c. First, we study
the influence of the reduction factor $\xi$ of antiproton-meson coupling
constants on the ACZ probability. To this aim, we have performed
additional calculations by choosing 
$\xi=0.15$  ($\mbox{Re}(V_{\rm opt})=-105$ MeV) 
and $\xi=1$ ($\mbox{Re}(V_{\rm opt})=-677$ MeV),
where the values of the Shr\"odinger equivalent potential at $E_{\rm lab}=0$ 
(see Ref. \cite{Larionov:2009tc} for details) in the centre of the $^{16}$O
nucleus are given in brackets. 
We recall that our default value $\xi=0.22$ ($\mbox{Re}(V_{\rm opt})=-153$ MeV)
is motivated by the best agreement of the GiBUU calculations \cite{Larionov:2009tc} 
with the measured antiproton absorption cross sections on nuclei. 
The value $\xi=0.15$ is in line with the results of $\bar p$-atomic X-ray transitions 
and radiochemical data analysis \cite{Friedman:2005ad}, while the extreme choice 
of $\xi=1$ corresponds to the G-parity transformed nucleon mean fields. 
As expected, larger (smaller) values of $\xi$ give rise to larger (smaller) ACZ 
probability at a given $\rho$.
Results are quite sensitive to the antiproton-meson coupling constants.
For instance, at $\rho=2\rho_0$, the ACZ probability is equal to zero in the 
case of $\xi=0.15$, i.e. the maximum nucleon density does not reach the value
of $2\rho_0$ at all with this value of $\xi$. At the same time, for $\xi=1$,
the ACZ probability is $\sim10^{-2}$ at $\rho=2\rho_0$ which is three orders
of magnitude larger than for $\xi=0.22$.

Finally, we discuss the sensitivity of our results to the choice of the
in-medium annihilation cross section related to the imaginary part of 
the antiproton optical potential 
\begin{equation}
   \mbox{Im}(V_{\rm opt})=-\frac{1}{2} < v_{\rm rel} \sigma_{\rm tot}^{\rm med} > \rho~.
                                \label{ImVopt}
\end{equation}
Here, $\sigma_{\rm tot}^{\rm med}$ is the total in-medium $\bar p$N cross section, which
includes both (in-medium) annihilative and Pauli-blocked nonannihilative contributions. 
Note, that in distinction to Eq. (\ref{Gamma_ann}) for the annihilation width,
Eq. (\ref{ImVopt}) contains the {\it total} $\bar p$N cross section.
Applying Eq. (\ref{ImVopt}) for the case of the HO formula for the annihilation 
probability per unit length (see Eq. (5) in Ref. \cite{Hernandez:1992rv})
leads to an extremely deep imaginary part of the antiproton optical potential,  
$\mbox{Im}(V_{\rm opt})\simeq-(1200-1700)$ MeV at the centre of the $^{16}$O nucleus,
with the uncertainty due to the antiproton mean field.
This is more than one order of magnitude larger than the value 
$\mbox{Im}(V_{\rm opt})\simeq-107$ MeV in our default choice of model parameters 
\cite{Larionov:2009tc}.

At this point, certainly, one wishes to get some phenomenological estimates 
of the antiproton-nucleus potential depths. Unfortunately, it is very hard to 
get stringent phenomenological constraints on the optical potential 
of a hadron at the nuclear centre \cite{Friedman:2007zz}.  
As it is known for a long time from pionic atoms and low-energy pion
nucleus scattering, the different density shapes of the potential
give the same result for the atomic and scattering data, while
they strongly differ at $\rho=\rho_0$ \cite{Seki:1983sh,Friedman:1983wz}.
In a similar way, the $\bar p$-nucleus scattering and absorption
cross sections (see \cite{Dalkarov:1986tu,Janouin:1986vh,Friedman:1986nf}
and refs. therein) and the $\bar p$-atomic data analysis 
\cite{Wong:1984fy,Friedman:2005ad,Friedman:2007zz} result in quite 
uncertain real and imaginary parts of the $\bar p$ optical potential 
at the nuclear centre $\mbox{Im}(V_{\rm opt})=-(70-150)$ MeV
and $\mbox{Re}(V_{\rm opt})=-(0-100)$ MeV. We stress that the actual
uncertainty in the potential depths may be much more due to
the extrapolation from far periphery of a nucleus using some assumed 
relation between the nuclear density and potential. 
Nevertheless, the known phenomenological values are in a fair agreement 
with our default model inputs. 

As one can see from Fig.~\ref{fig:probCompr_vs_rho}, 
using the HO formula leads to eight orders of magnitude smaller ACZ probability at 
$\rho=2\rho_0$ as compared to the calculation with the vacuum annihilation cross section 
(see the lines with filled boxes and with filled diamonds).
This is not surprising, since the $\rho$-dependent terms in Eq. (5) of Ref. 
\cite{Hernandez:1992rv} strongly enhance the annihilation rate at high densities. 
However, as discussed above, the mean field and phase-space effects should reduce 
the annihilation rate. 

Now we implement both effects simultaneously by introducing corresponding
multiplicative factors to the vacuum $\bar p$N annihilation
cross section. The resulting ACZ probability increases by eight orders of 
magnitude  with respect to the one given by the HO effect alone (see the lines with
upside-down filled triangles and with filled boxes in Fig.~\ref{fig:probCompr_vs_rho}).
i.e. practically brings it back to the original calculation with the vacuum annihilation 
cross section.
Certainly, this is only a rough estimate of the in-medium effects in the annihilation
cross section. In our opinion, the full calculation in the spirit of 
Refs.~\cite{Hernandez:1986ev,Hernandez:1989ij,Hernandez:1992rv}, but taking into account,
in-addition, the baryonic mean fields is needed in order to obtain the realistic values
for the antiproton width at high nucleon densities. 

%A better way would be to determine the value of the reduction factor $\xi$
%which produces the real and imaginary parts of the antiproton optical potential connected by the 
%dispersion relations, as it is done in Refs. \cite{Teis:1993ne,Teis:1994ie}. The selfconsistent
%determination of $\xi$, however, goes beyond the scope of the present work.

In spite of large uncertainties in the in-medium properties of antiproton,
we think that our standard choice of the model parameters, i.e. the vacuum
$\bar p$N annihilation cross section and the reduction factor $\xi=0.22$, 
is quite reasonable for the present study of compressional effects.   
As it has been shown within the GiBUU model in \cite{Larionov:2009tc}, this set
of parameters accounts for the $\bar p$ absorption data on nuclei at $p_{\rm lab} < 1$ GeV/c 
and pion and proton production data from $\bar p$ annihilation on nuclei at 608 MeV/c.

\section{Summary and outlook}
\label{Summary}

We have generalized our previous analysis of the nuclear compression dynamics
induced by an antiproton at rest \cite{Larionov:2008wy} to the case of a moving 
antiproton. The $\bar p$-nucleus collisions at the beam momenta of $0.3-10$ GeV/c
have been simulated within the transport GiBUU model \cite{GiBUU} with 
relativistic mean fields. 
In our two-stage calculational scheme, we apply, first, the standard parallel 
ensemble mode of GiBUU to determine the antibaryon coordinates and momenta 
at the annihilation point. We have studied in-detail the coordinate
and momentum distributions of annihilation points at different beam momenta. 
This calculation is performed in order to evaluate the probability that the
antibaryon has been slowed down and reached the nuclear interior before 
annihilation.
Those rare events which satisfy these conditions are used as the input for 
a more detailed calculation. Namely, we perform the coherent GiBUU runs 
\cite{Larionov:2008wy} initializing the antiproton at the given momentum 
and position inside the nucleus and following the evolution of the 
$\bar p$-nucleus system. In the coherent mode, the antibaryon-nucleon 
annihilation channels are switched off, but, instead, the survival probability
of the antiproton is determined as a function of time.
This allows to trace the compression process of the $\bar p$-nucleus system
in time and determine the probability of $\bar p$-annihilation in the compressed
nuclear configuration with the maximum nuclear density $\rho_{\rm max} \geq 2\rho_0$.

The results of our study are quite sensitive to the actual strengths of the real
and imaginary parts of the antiproton optical potential.
E.g., by choosing $\mbox{Re}(V_{\rm opt}(\rho_0))\simeq-100$ MeV instead of our
default $\mbox{Re}(V_{\rm opt}(\rho_0))\simeq-150$ MeV reduces the ACZ probability
by two orders of magnitude. The -100 MeV value of the real part of antiproton
optical potential is consistent with the X-ray data from antiprotonic atoms
and radiochemical data \cite{Friedman:2005ad}, while -150 MeV value is favoured
by GiBUU calculations of the antiproton absorption cross sections on nuclei
and of the pion and proton momentum spectra from $\bar p$ annihilation on nuclei
\cite{Larionov:2009tc}. Another important source of uncertainty is given by the
value of $\mbox{Im}(V_{\rm opt}(\rho_0))$. In our standard calculations,
we adopt $\mbox{Im}(V_{\rm opt}(\rho_0))\simeq-100$ MeV, which follows from a
simple $t\rho$-approximation and is consistent with 
$\mbox{Re}(V_{\rm opt}(\rho_0))\simeq-150$ MeV as shown in \cite{Larionov:2009tc}.
On the other hand, according to the model of Hern\'andez and Oset 
\cite{Hernandez:1986ev,Hernandez:1989ij,Hernandez:1992rv}, the antiproton
annihilation rate is increased by about one order of magnitude with respect to 
the simple $t \rho$-approximation, even at the normal nuclear matter density. 
This will result in $\mbox{Im}(V_{\rm opt}(\rho_0))\simeq-1500$ MeV, which is well
beyond the phenomenological expectations. If this were, indeed,
the case, the ACZ probability would be 5-8 orders of magnitude smaller 
than in our standard calculations. 

With all these reservations in mind, we now summarize the results of our 
standard calculations which use the phenomenological input parameters for 
the antiproton-nucleus interaction. In general, antiproton initializations 
in a central nuclear region with momenta 
of less or about the nucleon Fermi momentum lead to the maximum probability
of annihilation in the compressed zone of the order of $10^{-3}-10^{-1}$. 
The uncertainty is caused by unknown spatial spread of the antiproton distribution
function.
When combined with the actual antibaryon positions and momenta at the 
annihilation points determined from the first stage GiBUU simulation, this results in
the ACZ probability $\sim 10^{-5}-10^{-3}$ for the beam momenta of $3-10$ GeV/c.
We have found that, within this beam momentum range, the excitation function 
of the ACZ probability is very flat (c.f. Figs.~\ref{fig:sigCompr},
~\ref{fig:sigCompr_cuts}).
Therefore, the range $p_{\rm lab}=3-10$ GeV/c is quite well suited for the study 
of compressed nuclear systems. The beam momenta of about 1 GeV/c are clearly
disfavoured, since the antiproton is not decelerated enough due to the smallness
of the $\bar N N$ inelastic production cross section. At $p_{\rm lab} < 1$ GeV/c,
the ACZ probability grows up with decreasing beam momentum. However, additional
triggers demanding a fast proton \cite{PANDA} or large energy deposition 
\cite{Clover:1982qq,Cugnon:1986tx} are not very efficient for ACZ selection 
at small beam momenta. On the other hand, we have found, that these triggers
increase the ACZ probability by more than one order of magnitude in the beam
momentum range of 3-10 GeV/c. Such antiproton beams will be available at FAIR
which would be the ideal place to search for the nuclear compression
effects induced by antibaryons. By taking the expected luminosity 
$L=2\cdot10^{32}~\mbox{cm}^{-2}~\mbox{s}^{-1}$ for the $\overline{\rm P}$ANDA
experiment at FAIR \cite{PANDA}, the ACZ rate can be estimated as 
$Y = \sigma_{\rm compr} L \sim
10^2-10^3~\mbox{s}^{-1}$, where $\sigma_{\rm compr} \sim 10^{-3}-10^{-2}$ mb
is the ACZ cross section above 1 GeV/c (see Fig.~\ref{fig:sigCompr}).
Here, we would like to stress once again that, due to the presently 
not well known antiproton optical potential at $\rho \geq \rho_0$ and due to
uncertain spatial spread of the antiproton distribution function,
the above estimate of the ACZ rate has a rather large uncertainty.

We have also shown that the selection of small impact parameter events 
increases the ACZ probability by a factor of 2-3. This selection could 
be reached, e.g. by triggering on the events with a small azimuthal 
asymmetry of secondary particles. 

Some signals associated with the ACZ events have already been 
discussed in Refs. \cite{Mishustin:2004xa,Larionov:2008wy}. 
But, unfortunately, no unique signal suggested so far can alone 
be sufficient to identify the nuclear compression unambiguously.
Therefore, we believe that the combination of different signals,
e.g. emission of a fast proton plus large collective flow energy
of the nuclear fragments, would be a more promising strategy
to search for the ACZ events.  
Certainly, further theoretical studies are needed in order to find
the experimentally realizable ways to observe nuclear compression
in $\bar p$-nucleus collisions, in particular at FAIR energies.

\begin{acknowledgments}
We gratefully acknowledge the support by the Frankfurt Center for 
Scientific Computing. We thank O.~Buss, A.~Gillitzer, B.O.~Kerbikov,
U.~Mosel, J.~Ritman, and  I.A.~Pshenichnov for stimulating discussions.
 This work was partially supported by the Helmholtz International
Center for FAIR within the framework of the LOEWE program (Landesoffensive 
zur Entwicklung Wissenschaftlich-\"Okonomischer Exzellenz) launched by the 
State of Hesse, by the DFG Grant 436 RUS 113/957/0-1 (Germany), as well as
the Grants NS-7235.2010.2 and RFBR-09-02-91331 (Russia). 
\end{acknowledgments}
    
\bibliography{pbarDynCompr2}

%Merlin.mbs v4.21 2009-07-09.
\begin{thebibliography}{10}%
\makeatletter
\providecommand \@ifxundefined [1]{%
 \ifx #1\undefined \expandafter \@firstoftwo
 \else \expandafter \@secondoftwo
\fi
}%
\providecommand \@ifnum [1]{%
 \ifnum #1\expandafter \@firstoftwo
 \else \expandafter \@secondoftwo
\fi
}%
\providecommand \enquote [1]{``#1''}%
\providecommand \bibnamefont  [1]{#1}%
\providecommand \bibfnamefont [1]{#1}%
\providecommand \citenamefont [1]{#1}%
\providecommand\href[0]{\@sanitize\@href}%
\providecommand\@href[1]{\endgroup\@@startlink{#1}\endgroup\@@href}%
\providecommand\@@href[1]{#1\@@endlink}%
\providecommand \@sanitize [0]{\begingroup\catcode`\&12\catcode`\#12\relax}%
\@ifxundefined \pdfoutput {\@firstoftwo}{%
 \@ifnum{\z@=\pdfoutput}{\@firstoftwo}{\@secondoftwo}%
}{%
 \providecommand\@@startlink[1]{\leavevmode\special{html:<a href="#1">}}%
 \providecommand\@@endlink[0]{\special{html:</a>}}%
}{%
 \providecommand\@@startlink[1]{%
  \leavevmode
  \pdfstartlink
   attr{/Border[0 0 1 ]/H/I/C[0 1 1]}%
   user{/Subtype/Link/A<</Type/Action/S/URI/URI(#1)>>}%
  \relax
 }%
 \providecommand\@@endlink[0]{\pdfendlink}%
}%
\providecommand \url  [0]{\begingroup\@sanitize \@url }%
\providecommand \@url [1]{\endgroup\@href {#1}{\urlprefix}}%
\providecommand \urlprefix [0]{URL }%
\providecommand \Eprint[0]{\href }%
\@ifxundefined \urlstyle {%
  \providecommand \doi [1]{doi:\discretionary{}{}{}#1}%
}{%
  \providecommand \doi [0]{doi:\discretionary{}{}{}\begingroup
  \urlstyle{rm}\Url }%
}%
\providecommand \doibase [0]{http://dx.doi.org/}%
\providecommand \Doi[1]{\href{\doibase#1}}%
\providecommand \bibAnnote [3]{%
  \BibitemShut{#1}%
  \begin{quotation}\noindent
    \textsc{Key:}\ #2\\\textsc{Annotation:}\ #3%
  \end{quotation}%
}%
\providecommand \bibAnnoteFile [2]{%
  \IfFileExists{#2}{\bibAnnote {#1} {#2} {\input{#2}}}{}%
}%
\providecommand \typeout [0]{\immediate \write \m@ne }%
\providecommand \selectlanguage [0]{\@gobble}%
\providecommand \bibinfo [0]{\@secondoftwo}%
\providecommand \bibfield [0]{\@secondoftwo}%
\providecommand \translation [1]{[#1]}%
\providecommand \BibitemOpen[0]{}%
\providecommand \bibitemStop [0]{}%
\providecommand \bibitemNoStop [0]{.\EOS\space}%
\providecommand \EOS [0]{\spacefactor3000\relax}%
\providecommand \BibitemShut [1]{\csname bibitem#1\endcsname}%
%</preamble>
\bibitem{Collins:1974ky}%
  \BibitemOpen
  \bibfield{author}{%
  \bibinfo {author} {\bibfnamefont{J.~C.}\ \bibnamefont{Collins}}\ and\
  \bibinfo {author} {\bibfnamefont{M.~J.}\ \bibnamefont{Perry}},\ }%
  \bibfield{journal}{%
  \Doi{10.1103/PhysRevLett.34.1353}{\bibinfo {journal} {Phys. Rev. Lett.}}\ }%
  \textbf{\bibinfo {volume} {34}},\ \bibinfo {pages} {1353} (\bibinfo {year}
  {1975})%
  \bibAnnoteFile{NoStop}{Collins:1974ky}%
%%CITATION = PRLTA,34,1353;%%
\bibitem{Shuryak:1980tp}%
  \BibitemOpen
  \bibfield{author}{%
  \bibinfo {author} {\bibfnamefont{E.~V.}\ \bibnamefont{Shuryak}},\ }%
  \bibfield{journal}{%
  \Doi{10.1016/0370-1573(80)90105-2}{\bibinfo {journal} {Phys. Rept.}}\ }%
  \textbf{\bibinfo {volume} {61}},\ \bibinfo {pages} {71} (\bibinfo {year}
  {1980})%
  \bibAnnoteFile{NoStop}{Shuryak:1980tp}%
%%CITATION = PRPLC,61,71;%%
\bibitem{braun-munzinger:1031}%
  \BibitemOpen
  \bibfield{author}{%
  \bibinfo {author} {\bibfnamefont{P.}~\bibnamefont{Braun-Munzinger}}\ and\
  \bibinfo {author} {\bibfnamefont{J.}~\bibnamefont{Wambach}},\ }%
  \bibfield{journal}{%
  \Doi{10.1103/RevModPhys.81.1031}{\bibinfo {journal} {Rev. Mod. Phys.}}\ }%
  \textbf{\bibinfo {volume} {81}},\ \bibinfo {pages} {1031} (\bibinfo {year}
  {2009})%
  \bibAnnoteFile{NoStop}{braun-munzinger:1031}%
\bibitem{PhysRevLett.36.88}%
  \BibitemOpen
  \bibfield{author}{%
  \bibinfo {author} {\bibfnamefont{J.}~\bibnamefont{Hofmann}}, \bibinfo
  {author} {\bibfnamefont{H.}~\bibnamefont{St\"ocker}}, \bibinfo {author}
  {\bibfnamefont{U.}~\bibnamefont{Heinz}}, \bibinfo {author}
  {\bibfnamefont{W.}~\bibnamefont{Scheid}},\ and\ \bibinfo {author}
  {\bibfnamefont{W.}~\bibnamefont{Greiner}},\ }%
  \bibfield{journal}{%
  \Doi{10.1103/PhysRevLett.36.88}{\bibinfo {journal} {Phys. Rev. Lett.}}\ }%
  \textbf{\bibinfo {volume} {36}},\ \bibinfo {pages} {88} (\bibinfo {year}
  {1976})%
  \bibAnnoteFile{NoStop}{PhysRevLett.36.88}%
\bibitem{PhysRevLett.86.1982}%
  \BibitemOpen
  \bibfield{author}{%
  \bibinfo {author} {\bibfnamefont{K.}~\bibnamefont{Tanida}}, \bibinfo {author}
  {\bibfnamefont{H.}~\bibnamefont{Tamura}}, \bibinfo {author}
  {\bibfnamefont{D.}~\bibnamefont{Abe}}, \bibinfo {author}
  {\bibfnamefont{H.}~\bibnamefont{Akikawa}}, \bibinfo {author}
  {\bibfnamefont{K.}~\bibnamefont{Araki}}, \bibinfo {author}
  {\bibfnamefont{H.}~\bibnamefont{Bhang}}, \bibinfo {author}
  {\bibfnamefont{T.}~\bibnamefont{Endo}}, \bibinfo {author}
  {\bibfnamefont{Y.}~\bibnamefont{Fujii}}, \bibinfo {author}
  {\bibfnamefont{T.}~\bibnamefont{Fukuda}}, \bibinfo {author}
  {\bibfnamefont{O.}~\bibnamefont{Hashimoto}}, \bibinfo {author}
  {\bibfnamefont{K.}~\bibnamefont{Imai}}, \bibinfo {author}
  {\bibfnamefont{H.}~\bibnamefont{Hotchi}}, \bibinfo {author}
  {\bibfnamefont{Y.}~\bibnamefont{Kakiguchi}}, \bibinfo {author}
  {\bibfnamefont{J.~H.}\ \bibnamefont{Kim}}, \bibinfo {author}
  {\bibfnamefont{Y.~D.}\ \bibnamefont{Kim}}, \bibinfo {author}
  {\bibfnamefont{T.}~\bibnamefont{Miyoshi}}, \bibinfo {author}
  {\bibfnamefont{T.}~\bibnamefont{Murakami}}, \bibinfo {author}
  {\bibfnamefont{T.}~\bibnamefont{Nagae}}, \bibinfo {author}
  {\bibfnamefont{H.}~\bibnamefont{Noumi}}, \bibinfo {author}
  {\bibfnamefont{H.}~\bibnamefont{Outa}}, \bibinfo {author}
  {\bibfnamefont{K.}~\bibnamefont{Ozawa}}, \bibinfo {author}
  {\bibfnamefont{T.}~\bibnamefont{Saito}}, \bibinfo {author}
  {\bibfnamefont{J.}~\bibnamefont{Sasao}}, \bibinfo {author}
  {\bibfnamefont{Y.}~\bibnamefont{Sato}}, \bibinfo {author}
  {\bibfnamefont{S.}~\bibnamefont{Satoh}}, \bibinfo {author}
  {\bibfnamefont{R.~I.}\ \bibnamefont{Sawafta}},\ and\ \bibinfo {author}
  {\bibfnamefont{M.}~\bibnamefont{Sekimoto}},\ }%
  \bibfield{journal}{%
  \Doi{10.1103/PhysRevLett.86.1982}{\bibinfo {journal} {Phys. Rev. Lett.}}\ }%
  \textbf{\bibinfo {volume} {86}},\ \bibinfo {pages} {1982} (\bibinfo {month}
  {Mar}\ \bibinfo {year} {2001})%
  \bibAnnoteFile{NoStop}{PhysRevLett.86.1982}%
\bibitem{Schaffner:1993nn}%
  \BibitemOpen
  \bibfield{author}{%
  \bibinfo {author} {\bibfnamefont{J.}~\bibnamefont{Schaffner}}, \bibinfo
  {author} {\bibfnamefont{C.~B.}\ \bibnamefont{Dover}}, \bibinfo {author}
  {\bibfnamefont{A.}~\bibnamefont{Gal}}, \bibinfo {author}
  {\bibfnamefont{C.}~\bibnamefont{Greiner}},\ and\ \bibinfo {author}
  {\bibfnamefont{H.}~\bibnamefont{St\"ocker}},\ }%
  \bibfield{journal}{%
  \Doi{10.1103/PhysRevLett.71.1328}{\bibinfo {journal} {Phys. Rev. Lett.}}\ }%
  \textbf{\bibinfo {volume} {71}},\ \bibinfo {pages} {1328} (\bibinfo {year}
  {1993})%
  \bibAnnoteFile{NoStop}{Schaffner:1993nn}%
%%CITATION = PRLTA,71,1328;%%
\bibitem{Schaffner199435}%
  \BibitemOpen
  \bibfield{author}{%
  \bibinfo {author} {\bibfnamefont{J.}~\bibnamefont{Schaffner}}, \bibinfo
  {author} {\bibfnamefont{C.~B.}\ \bibnamefont{Dover}}, \bibinfo {author}
  {\bibfnamefont{A.}~\bibnamefont{Gal}}, \bibinfo {author}
  {\bibfnamefont{C.}~\bibnamefont{Greiner}}, \bibinfo {author}
  {\bibfnamefont{D.~J.}\ \bibnamefont{Millener}},\ and\ \bibinfo {author}
  {\bibfnamefont{H.}~\bibnamefont{St\"ocker}},\ }%
  \bibfield{journal}{%
  \Doi{DOI: 10.1006/aphy.1994.1090}{\bibinfo {journal} {Ann. Phys.}}\ }%
  \textbf{\bibinfo {volume} {235}},\ \bibinfo {pages} {35 } (\bibinfo {year}
  {1994})%
  \bibAnnoteFile{NoStop}{Schaffner199435}%
\bibitem{Akaishi:2002bg}%
  \BibitemOpen
  \bibfield{author}{%
  \bibinfo {author} {\bibfnamefont{Y.}~\bibnamefont{Akaishi}}\ and\ \bibinfo
  {author} {\bibfnamefont{T.}~\bibnamefont{Yamazaki}},\ }%
  \bibfield{journal}{%
  \Doi{10.1103/PhysRevC.65.044005}{\bibinfo {journal} {Phys. Rev.}}\ }%
  \textbf{\bibinfo {volume} {C65}},\ \bibinfo {pages} {044005} (\bibinfo {year}
  {2002})%
  \bibAnnoteFile{NoStop}{Akaishi:2002bg}%
%%CITATION = PHRVA,C65,044005;%%
\bibitem{Hayano:2008zzd}%
  \BibitemOpen
  \bibfield{author}{%
  \bibinfo {author} {\bibfnamefont{R.~S.}\ \bibnamefont{Hayano}} \emph{et~al.}
  (\bibinfo {collaboration} {KEK-E570}),\ }%
  \bibfield{journal}{%
  \Doi{10.1142/S021773230802968X}{\bibinfo {journal} {Mod. Phys. Lett.}}\ }%
  \textbf{\bibinfo {volume} {A23}},\ \bibinfo {pages} {2505} (\bibinfo {year}
  {2008})%
  \bibAnnoteFile{NoStop}{Hayano:2008zzd}%
%%CITATION = MPLAE,A23,2505;%%
\bibitem{Mares:2006vk}%
  \BibitemOpen
  \bibfield{author}{%
  \bibinfo {author} {\bibfnamefont{J.}~\bibnamefont{Mares}}, \bibinfo {author}
  {\bibfnamefont{E.}~\bibnamefont{Friedman}},\ and\ \bibinfo {author}
  {\bibfnamefont{A.}~\bibnamefont{Gal}},\ }%
  \bibfield{journal}{%
  \Doi{10.1016/j.nuclphysa.2006.02.010}{\bibinfo {journal} {Nucl. Phys.}}\ }%
  \textbf{\bibinfo {volume} {A770}},\ \bibinfo {pages} {84} (\bibinfo {year}
  {2006}),\
  \Eprint{http://arxiv.org/abs/nucl-th/0601009}{arXiv:nucl-th/0601009}%
  \bibAnnoteFile{NoStop}{Mares:2006vk}%
%%CITATION = NUCL-TH/0601009;%%
\bibitem{Ramos:1999ku}%
  \BibitemOpen
  \bibfield{author}{%
  \bibinfo {author} {\bibfnamefont{A.}~\bibnamefont{Ramos}}\ and\ \bibinfo
  {author} {\bibfnamefont{E.}~\bibnamefont{Oset}},\ }%
  \bibfield{journal}{%
  \Doi{10.1016/S0375-9474(99)00846-5}{\bibinfo {journal} {Nucl. Phys.}}\ }%
  \textbf{\bibinfo {volume} {A671}},\ \bibinfo {pages} {481} (\bibinfo {year}
  {2000}),\
  \Eprint{http://arxiv.org/abs/nucl-th/9906016}{arXiv:nucl-th/9906016}%
  \bibAnnoteFile{NoStop}{Ramos:1999ku}%
%%CITATION = NUCL-TH/9906016;%%
\bibitem{Oset:2005sn}%
  \BibitemOpen
  \bibfield{author}{%
  \bibinfo {author} {\bibfnamefont{E.}~\bibnamefont{Oset}}\ and\ \bibinfo
  {author} {\bibfnamefont{H.}~\bibnamefont{Toki}},\ }%
  \bibfield{journal}{%
  \Doi{10.1103/PhysRevC.74.015207}{\bibinfo {journal} {Phys. Rev.}}\ }%
  \textbf{\bibinfo {volume} {C74}},\ \bibinfo {pages} {015207} (\bibinfo {year}
  {2006})%
  \bibAnnoteFile{NoStop}{Oset:2005sn}%
%%CITATION = NUCL-TH/0509048;%%
\bibitem{Hyodo:2007jq}%
  \BibitemOpen
  \bibfield{author}{%
  \bibinfo {author} {\bibfnamefont{T.}~\bibnamefont{Hyodo}}\ and\ \bibinfo
  {author} {\bibfnamefont{W.}~\bibnamefont{Weise}},\ }%
  \bibfield{journal}{%
  \Doi{10.1103/PhysRevC.77.035204}{\bibinfo {journal} {Phys. Rev.}}\ }%
  \textbf{\bibinfo {volume} {C77}},\ \bibinfo {pages} {035204} (\bibinfo {year}
  {2008})%
  \bibAnnoteFile{NoStop}{Hyodo:2007jq}%
%%CITATION = 0712.1613;%%
\bibitem{Dote:2008hw}%
  \BibitemOpen
  \bibfield{author}{%
  \bibinfo {author} {\bibfnamefont{A.}~\bibnamefont{Dote}}, \bibinfo {author}
  {\bibfnamefont{T.}~\bibnamefont{Hyodo}},\ and\ \bibinfo {author}
  {\bibfnamefont{W.}~\bibnamefont{Weise}},\ }%
  \bibfield{journal}{%
  \Doi{10.1103/PhysRevC.79.014003}{\bibinfo {journal} {Phys. Rev.}}\ }%
  \textbf{\bibinfo {volume} {C79}},\ \bibinfo {pages} {014003} (\bibinfo {year}
  {2009})%
  \bibAnnoteFile{NoStop}{Dote:2008hw}%
%%CITATION = 0806.4917;%%
\bibitem{Akaishi:2010wt}%
  \BibitemOpen
  \bibfield{author}{%
  \bibinfo {author} {\bibfnamefont{Y.}~\bibnamefont{Akaishi}}, \bibinfo
  {author} {\bibfnamefont{T.}~\bibnamefont{Yamazaki}}, \bibinfo {author}
  {\bibfnamefont{M.}~\bibnamefont{Obu}},\ and\ \bibinfo {author}
  {\bibfnamefont{M.}~\bibnamefont{Wada}},\ }%
  \bibfield{journal}{%
  \Doi{10.1016/j.nuclphysa.2010.01.176}{\bibinfo {journal} {Nucl. Phys.}}\ }%
  \textbf{\bibinfo {volume} {A835}},\ \bibinfo {pages} {67} (\bibinfo {year}
  {2010})%
  \bibAnnoteFile{NoStop}{Akaishi:2010wt}%
%%CITATION = 1002.2560;%%
\bibitem{Yamazaki:2010mu}%
  \BibitemOpen
  \bibfield{author}{%
  \bibinfo {author} {\bibfnamefont{T.}~\bibnamefont{Yamazaki}} \emph{et~al.},\
  }%
  \bibfield{journal}{%
  \Doi{10.1103/PhysRevLett.104.132502}{\bibinfo {journal} {Phys. Rev. Lett.}}\
  }%
  \textbf{\bibinfo {volume} {104}},\ \bibinfo {pages} {132502} (\bibinfo {year}
  {2010})%
  \bibAnnoteFile{NoStop}{Yamazaki:2010mu}%
%%CITATION = 1002.3526;%%
\bibitem{Buervenich:2002ns}%
  \BibitemOpen
  \bibfield{author}{%
  \bibinfo {author} {\bibfnamefont{T.~J.}\ \bibnamefont{B\"urvenich}}, \bibinfo
  {author} {\bibfnamefont{I.~N.}\ \bibnamefont{Mishustin}}, \bibinfo {author}
  {\bibfnamefont{L.~M.}\ \bibnamefont{Satarov}}, \bibinfo {author}
  {\bibfnamefont{J.~A.}\ \bibnamefont{Maruhn}}, \bibinfo {author}
  {\bibfnamefont{H.}~\bibnamefont{St\"ocker}},\ and\ \bibinfo {author}
  {\bibfnamefont{W.}~\bibnamefont{Greiner}},\ }%
  \bibfield{journal}{%
  \Doi{10.1016/S0370-2693(02)02351-1}{\bibinfo {journal} {Phys. Lett.}}\ }%
  \textbf{\bibinfo {volume} {B542}},\ \bibinfo {pages} {261} (\bibinfo {year}
  {2002})%
  \bibAnnoteFile{NoStop}{Buervenich:2002ns}%
%%CITATION = NUCL-TH/0207011;%%
\bibitem{Mishustin:2004xa}%
  \BibitemOpen
  \bibfield{author}{%
  \bibinfo {author} {\bibfnamefont{I.~N.}\ \bibnamefont{Mishustin}}, \bibinfo
  {author} {\bibfnamefont{L.~M.}\ \bibnamefont{Satarov}}, \bibinfo {author}
  {\bibfnamefont{T.~J.}\ \bibnamefont{B\"urvenich}}, \bibinfo {author}
  {\bibfnamefont{H.}~\bibnamefont{St\"ocker}},\ and\ \bibinfo {author}
  {\bibfnamefont{W.}~\bibnamefont{Greiner}},\ }%
  \bibfield{journal}{%
  \Doi{10.1103/PhysRevC.71.035201}{\bibinfo {journal} {Phys. Rev.}}\ }%
  \textbf{\bibinfo {volume} {C71}},\ \bibinfo {pages} {035201} (\bibinfo {year}
  {2005})%
  \bibAnnoteFile{NoStop}{Mishustin:2004xa}%
%%CITATION = NUCL-TH/0404026;%%
\bibitem{Friedman:2005ad}%
  \BibitemOpen
  \bibfield{author}{%
  \bibinfo {author} {\bibnamefont{{Friedman, E. and Gal, A. and Mare\v{s},
  J.}}},\ }%
  \bibfield{journal}{%
  \Doi{10.1016/j.nuclphysa.2005.08.001}{\bibinfo {journal} {Nucl. Phys.}}\ }%
  \textbf{\bibinfo {volume} {A761}},\ \bibinfo {pages} {283} (\bibinfo {year}
  {2005})%
  \bibAnnoteFile{NoStop}{Friedman:2005ad}%
%%CITATION = NUCL-TH/0504030;%%
\bibitem{Pochodzalla:2008ju}%
  \BibitemOpen
  \bibfield{author}{%
  \bibinfo {author} {\bibfnamefont{J.}~\bibnamefont{Pochodzalla}},\ }%
  \bibfield{journal}{%
  \Doi{10.1016/j.physletb.2008.10.016}{\bibinfo {journal} {Phys. Lett.}}\ }%
  \textbf{\bibinfo {volume} {B669}},\ \bibinfo {pages} {306} (\bibinfo {year}
  {2008})%
  \bibAnnoteFile{NoStop}{Pochodzalla:2008ju}%
%%CITATION = 0807.3302;%%
\bibitem{PANDA}%
  \BibitemOpen
  \bibfield{author}{%
  \bibinfo {author} {\bibnamefont{\mbox{The~PANDA~Collaboration}}}, \bibinfo
  {author} {\bibfnamefont{M.~F.~M.}\ \bibnamefont{Lutz}}, \bibinfo {author}
  {\bibfnamefont{B.}~\bibnamefont{Pire}}, \bibinfo {author}
  {\bibfnamefont{O.}~\bibnamefont{Scholten}},\ and\ \bibinfo {author}
  {\bibfnamefont{R.}~\bibnamefont{Timmermans}},\ }%
  \bibfield{journal}{%
  \bibinfo {journal} {arXiv:0903.3905}}%
   (\bibinfo {year} {2009})%
  \bibAnnoteFile{NoStop}{PANDA}%
\bibitem{Larionov:2009tc}%
  \BibitemOpen
  \bibfield{author}{%
  \bibinfo {author} {\bibfnamefont{A.~B.}\ \bibnamefont{Larionov}}, \bibinfo
  {author} {\bibfnamefont{I.~A.}\ \bibnamefont{Pshenichnov}}, \bibinfo {author}
  {\bibfnamefont{I.~N.}\ \bibnamefont{Mishustin}},\ and\ \bibinfo {author}
  {\bibfnamefont{W.}~\bibnamefont{Greiner}},\ }%
  \bibfield{journal}{%
  \Doi{10.1103/PhysRevC.80.021601}{\bibinfo {journal} {Phys. Rev.}}\ }%
  \textbf{\bibinfo {volume} {C80}},\ \bibinfo {pages} {021601} (\bibinfo {year}
  {2009})%
  \bibAnnoteFile{NoStop}{Larionov:2009tc}%
%%CITATION = 0903.2152;%%
\bibitem{Hernandez:1986ev}%
  \BibitemOpen
  \bibfield{author}{%
  \bibinfo {author} {\bibfnamefont{E.}~\bibnamefont{Hern\'andez}}\ and\
  \bibinfo {author} {\bibfnamefont{E.}~\bibnamefont{Oset}},\ }%
  \bibfield{journal}{%
  \Doi{10.1016/0370-2693(86)90033-X}{\bibinfo {journal} {Phys. Lett.}}\ }%
  \textbf{\bibinfo {volume} {B181}},\ \bibinfo {pages} {207} (\bibinfo {year}
  {1986})%
  \bibAnnoteFile{NoStop}{Hernandez:1986ev}%
%%CITATION = PHLTA,B181,207;%%
\bibitem{Hernandez:1989ij}%
  \BibitemOpen
  \bibfield{author}{%
  \bibinfo {author} {\bibfnamefont{E.}~\bibnamefont{Hern\'andez}}\ and\
  \bibinfo {author} {\bibfnamefont{E.}~\bibnamefont{Oset}},\ }%
  \bibfield{journal}{%
  \Doi{10.1016/0375-9474(89)90097-3}{\bibinfo {journal} {Nucl. Phys.}}\ }%
  \textbf{\bibinfo {volume} {A493}},\ \bibinfo {pages} {453} (\bibinfo {year}
  {1989})%
  \bibAnnoteFile{NoStop}{Hernandez:1989ij}%
%%CITATION = NUPHA,A493,453;%%
\bibitem{Hernandez:1992rv}%
  \BibitemOpen
  \bibfield{author}{%
  \bibinfo {author} {\bibfnamefont{E.}~\bibnamefont{Hern\'andez}}\ and\
  \bibinfo {author} {\bibfnamefont{E.}~\bibnamefont{Oset}},\ }%
  \bibfield{journal}{%
  \Doi{10.1007/BF01298481}{\bibinfo {journal} {Z. Phys.}}\ }%
  \textbf{\bibinfo {volume} {A341}},\ \bibinfo {pages} {201} (\bibinfo {year}
  {1992})%
  \bibAnnoteFile{NoStop}{Hernandez:1992rv}%
%%CITATION = ZEPYA,A341,201;%%
\bibitem{Larionov:2008wy}%
  \BibitemOpen
  \bibfield{author}{%
  \bibinfo {author} {\bibfnamefont{A.~B.}\ \bibnamefont{Larionov}}, \bibinfo
  {author} {\bibfnamefont{I.~N.}\ \bibnamefont{Mishustin}}, \bibinfo {author}
  {\bibfnamefont{L.~M.}\ \bibnamefont{Satarov}},\ and\ \bibinfo {author}
  {\bibfnamefont{W.}~\bibnamefont{Greiner}},\ }%
  \bibfield{journal}{%
  \Doi{10.1103/PhysRevC.78.014604}{\bibinfo {journal} {Phys. Rev.}}\ }%
  \textbf{\bibinfo {volume} {C78}},\ \bibinfo {pages} {014604} (\bibinfo {year}
  {2008})%
  \bibAnnoteFile{NoStop}{Larionov:2008wy}%
%%CITATION =0802.1845;%%
\bibitem{GiBUU}%
  \BibitemOpen
  \url{http://gibuu.physik.uni-giessen.de/GiBUU}%
  \bibAnnoteFile{NoStop}{GiBUU}%
\bibitem{Larionov:2007hy}%
  \BibitemOpen
  \bibfield{author}{%
  \bibinfo {author} {\bibfnamefont{A.~B.}\ \bibnamefont{Larionov}}, \bibinfo
  {author} {\bibfnamefont{O.}~\bibnamefont{Buss}}, \bibinfo {author}
  {\bibfnamefont{K.}~\bibnamefont{Gallmeister}},\ and\ \bibinfo {author}
  {\bibfnamefont{U.}~\bibnamefont{Mosel}},\ }%
  \bibfield{journal}{%
  \Doi{10.1103/PhysRevC.76.044909}{\bibinfo {journal} {Phys. Rev.}}\ }%
  \textbf{\bibinfo {volume} {C76}},\ \bibinfo {pages} {044909} (\bibinfo {year}
  {2007})%
  \bibAnnoteFile{NoStop}{Larionov:2007hy}%
%%CITATION = 0704.1785;%%
\bibitem{Gaitanos:2007mm}%
  \BibitemOpen
  \bibfield{author}{%
  \bibinfo {author} {\bibfnamefont{T.}~\bibnamefont{Gaitanos}}, \bibinfo
  {author} {\bibfnamefont{H.}~\bibnamefont{Lenske}},\ and\ \bibinfo {author}
  {\bibfnamefont{U.}~\bibnamefont{Mosel}},\ }%
  \bibfield{journal}{%
  \Doi{10.1016/j.physletb.2008.04.011}{\bibinfo {journal} {Phys. Lett.}}\ }%
  \textbf{\bibinfo {volume} {B663}},\ \bibinfo {pages} {197} (\bibinfo {year}
  {2008})%
  \bibAnnoteFile{NoStop}{Gaitanos:2007mm}%
%%CITATION = 0712.3292;%%
\bibitem{Lalazissis:1996rd}%
  \BibitemOpen
  \bibfield{author}{%
  \bibinfo {author} {\bibnamefont{{Lalazissis, G. A. and K\"onig, J. and Ring,
  P.}}},\ }%
  \bibfield{journal}{%
  \Doi{10.1103/PhysRevC.55.540}{\bibinfo {journal} {Phys. Rev.}}\ }%
  \textbf{\bibinfo {volume} {C55}},\ \bibinfo {pages} {540} (\bibinfo {year}
  {1997})%
  \bibAnnoteFile{NoStop}{Lalazissis:1996rd}%
%%CITATION = NUCL-TH/9607039;%%
\bibitem{Aichelin:1991xy}%
  \BibitemOpen
  \bibfield{author}{%
  \bibinfo {author} {\bibfnamefont{J.}~\bibnamefont{Aichelin}},\ }%
  \bibfield{journal}{%
  \Doi{10.1016/0370-1573(91)90094-3}{\bibinfo {journal} {Phys. Rept.}}\ }%
  \textbf{\bibinfo {volume} {202}},\ \bibinfo {pages} {233} (\bibinfo {year}
  {1991})%
  \bibAnnoteFile{NoStop}{Aichelin:1991xy}%
%%CITATION = PRPLC,202,233;%%
\bibitem{Dote:2005nb}%
  \BibitemOpen
  \bibfield{author}{%
  \bibinfo {author} {\bibfnamefont{A.}~\bibnamefont{Dote}}, \bibinfo {author}
  {\bibfnamefont{Y.}~\bibnamefont{Akaishi}},\ and\ \bibinfo {author}
  {\bibfnamefont{T.}~\bibnamefont{Yamazaki}},\ }%
  \bibfield{journal}{%
  \Doi{10.1016/j.nuclphysa.2005.01.029}{\bibinfo {journal} {Nucl. Phys.}}\ }%
  \textbf{\bibinfo {volume} {A754}},\ \bibinfo {pages} {391} (\bibinfo {year}
  {2005})%
  \bibAnnoteFile{NoStop}{Dote:2005nb}%
%%CITATION = NUPHA,A754,391;%%
\bibitem{Mishustin:2008ch}%
  \BibitemOpen
  \bibfield{author}{%
  \bibinfo {author} {\bibfnamefont{I.~N.}\ \bibnamefont{Mishustin}}\ and\
  \bibinfo {author} {\bibfnamefont{A.~B.}\ \bibnamefont{Larionov}},\ }%
  \bibfield{journal}{%
  \bibinfo {journal} {Hyperfine Interact.}\ }%
  \textbf{\bibinfo {volume} {194}},\ \bibinfo {pages} {263} (\bibinfo {year}
  {2009})%
  \bibAnnoteFile{NoStop}{Mishustin:2008ch}%
%%CITATION = 0810.4030;%%
\bibitem{Montanet:1994xu}%
  \BibitemOpen
  \bibfield{author}{%
  \bibinfo {author} {\bibfnamefont{L.}~\bibnamefont{Montanet}} \emph{et~al.}
  (\bibinfo {collaboration} {Particle Data Group}),\ }%
  \bibfield{journal}{%
  \Doi{10.1103/PhysRevD.50.1173}{\bibinfo {journal} {Phys. Rev.}}\ }%
  \textbf{\bibinfo {volume} {D50}},\ \bibinfo {pages} {1173} (\bibinfo {year}
  {1994})%
  \bibAnnoteFile{NoStop}{Montanet:1994xu}%
%%CITATION = PHRVA,D50,1173;%%
\bibitem{Rafelski:1979nt}%
  \BibitemOpen
  \bibfield{author}{%
  \bibinfo {author} {\bibfnamefont{J.}~\bibnamefont{Rafelski}},\ }%
  \bibfield{journal}{%
  \Doi{10.1016/0370-2693(80)90450-5}{\bibinfo {journal} {Phys. Lett.}}\ }%
  \textbf{\bibinfo {volume} {B91}},\ \bibinfo {pages} {281} (\bibinfo {year}
  {1980})%
  \bibAnnoteFile{NoStop}{Rafelski:1979nt}%
%%CITATION = PHLTA,B91,281;%%
\bibitem{Salvini:2005sb}%
  \BibitemOpen
  \bibfield{author}{%
  \bibinfo {author} {\bibfnamefont{P.}~\bibnamefont{Salvini}}, \bibinfo
  {author} {\bibfnamefont{A.}~\bibnamefont{Panzarasa}},\ and\ \bibinfo {author}
  {\bibfnamefont{G.}~\bibnamefont{Bendiscioli}},\ }%
  \bibfield{journal}{%
  \Doi{10.1016/j.nuclphysa.2005.06.009}{\bibinfo {journal} {Nucl. Phys.}}\ }%
  \textbf{\bibinfo {volume} {A760}},\ \bibinfo {pages} {349} (\bibinfo {year}
  {2005})%
  \bibAnnoteFile{NoStop}{Salvini:2005sb}%
%%CITATION = NUPHA,A760,349;%%
\bibitem{Bendiscioli:2009zza}%
  \BibitemOpen
  \bibfield{author}{%
  \bibinfo {author} {\bibfnamefont{G.}~\bibnamefont{Bendiscioli}}, \bibinfo
  {author} {\bibfnamefont{T.}~\bibnamefont{Bressani}}, \bibinfo {author}
  {\bibfnamefont{L.}~\bibnamefont{Lavezzi}}, \bibinfo {author}
  {\bibfnamefont{A.}~\bibnamefont{Panzarasa}},\ and\ \bibinfo {author}
  {\bibfnamefont{P.}~\bibnamefont{Salvini}},\ }%
  \bibfield{journal}{%
  \Doi{10.1016/j.nuclphysa.2008.10.013}{\bibinfo {journal} {Nucl. Phys.}}\ }%
  \textbf{\bibinfo {volume} {A815}},\ \bibinfo {pages} {67} (\bibinfo {year}
  {2009})%
  \bibAnnoteFile{NoStop}{Bendiscioli:2009zza}%
%%CITATION = NUPHA,A815,67;%%
\bibitem{Stoecker:1986ci}%
  \BibitemOpen
  \bibfield{author}{%
  \bibinfo {author} {\bibfnamefont{H.}~\bibnamefont{St\"ocker}}\ and\ \bibinfo
  {author} {\bibfnamefont{W.}~\bibnamefont{Greiner}},\ }%
  \bibfield{journal}{%
  \Doi{10.1016/0370-1573(86)90131-6}{\bibinfo {journal} {Phys. Rept.}}\ }%
  \textbf{\bibinfo {volume} {137}},\ \bibinfo {pages} {277} (\bibinfo {year}
  {1986})%
  \bibAnnoteFile{NoStop}{Stoecker:1986ci}%
%%CITATION = PRPLC,137,277;%%
\bibitem{Cugnon:1989}%
  \BibitemOpen
  \bibfield{author}{%
  \bibinfo {author} {\bibfnamefont{J.}~\bibnamefont{Cugnon}}\ and\ \bibinfo
  {author} {\bibfnamefont{J.}~\bibnamefont{Vandermeulen}},\ }%
  \bibfield{journal}{%
  \bibinfo {journal} {Ann. Phys. (France)}\ }%
  \textbf{\bibinfo {volume} {14}},\ \bibinfo {pages} {49} (\bibinfo {year}
  {1989})%
  \bibAnnoteFile{NoStop}{Cugnon:1989}%
\bibitem{Clover:1982qq}%
  \BibitemOpen
  \bibfield{author}{%
  \bibinfo {author} {\bibfnamefont{M.~R.}\ \bibnamefont{Clover}}, \bibinfo
  {author} {\bibfnamefont{R.~M.}\ \bibnamefont{DeVries}}, \bibinfo {author}
  {\bibfnamefont{N.~J.}\ \bibnamefont{DiGiacomo}},\ and\ \bibinfo {author}
  {\bibfnamefont{Y.}~\bibnamefont{Yariv}},\ }%
  \bibfield{journal}{%
  \Doi{10.1103/PhysRevC.26.2138}{\bibinfo {journal} {Phys. Rev.}}\ }%
  \textbf{\bibinfo {volume} {C26}},\ \bibinfo {pages} {2138} (\bibinfo {year}
  {1982})%
  \bibAnnoteFile{NoStop}{Clover:1982qq}%
%%CITATION = PHRVA,C26,2138;%%
\bibitem{Cugnon:1986tx}%
  \BibitemOpen
  \bibfield{author}{%
  \bibinfo {author} {\bibfnamefont{J.}~\bibnamefont{Cugnon}}\ and\ \bibinfo
  {author} {\bibfnamefont{J.}~\bibnamefont{Vandermeulen}},\ }%
  \bibfield{journal}{%
  \Doi{10.1016/0375-9474(85)90568-8}{\bibinfo {journal} {Nucl. Phys.}}\ }%
  \textbf{\bibinfo {volume} {A445}},\ \bibinfo {pages} {717} (\bibinfo {year}
  {1985})%
  \bibAnnoteFile{NoStop}{Cugnon:1986tx}%
%%CITATION = NUPHA,A445,717;%%
\bibitem{Cugnon:1984zp}%
  \BibitemOpen
  \bibfield{author}{%
  \bibinfo {author} {\bibfnamefont{J.}~\bibnamefont{Cugnon}}\ and\ \bibinfo
  {author} {\bibfnamefont{J.}~\bibnamefont{Vandermeulen}},\ }%
  \bibfield{journal}{%
  \Doi{10.1016/0370-2693(84)90633-6}{\bibinfo {journal} {Phys. Lett.}}\ }%
  \textbf{\bibinfo {volume} {B146}},\ \bibinfo {pages} {16} (\bibinfo {year}
  {1984})%
  \bibAnnoteFile{NoStop}{Cugnon:1984zp}%
%%CITATION = PHLTA,B146,16;%%
\bibitem{Cugnon:1989hj}%
  \BibitemOpen
  \bibfield{author}{%
  \bibinfo {author} {\bibfnamefont{J.}~\bibnamefont{Cugnon}}\ and\ \bibinfo
  {author} {\bibfnamefont{J.}~\bibnamefont{Vandermeulen}},\ }%
  \bibfield{journal}{%
  \Doi{10.1103/PhysRevC.39.181}{\bibinfo {journal} {Phys. Rev.}}\ }%
  \textbf{\bibinfo {volume} {C39}},\ \bibinfo {pages} {181} (\bibinfo {year}
  {1989})%
  \bibAnnoteFile{NoStop}{Cugnon:1989hj}%
%%CITATION = PHRVA,C39,181;%%
\bibitem{Kahana:1992hx}%
  \BibitemOpen
  \bibfield{author}{%
  \bibinfo {author} {\bibfnamefont{S.~H.}\ \bibnamefont{Kahana}}, \bibinfo
  {author} {\bibfnamefont{Y.}~\bibnamefont{Pang}}, \bibinfo {author}
  {\bibfnamefont{T.}~\bibnamefont{Schlagel}},\ and\ \bibinfo {author}
  {\bibfnamefont{C.~B.}\ \bibnamefont{Dover}},\ }%
  \bibfield{journal}{%
  \Doi{10.1103/PhysRevC.47.R1356}{\bibinfo {journal} {Phys. Rev.}}\ }%
  \textbf{\bibinfo {volume} {C47}},\ \bibinfo {pages} {1356} (\bibinfo {year}
  {1993})%
  \bibAnnoteFile{NoStop}{Kahana:1992hx}%
%%CITATION = PHRVA,C47,1356;%%
\bibitem{Pang:1996ez}%
  \BibitemOpen
  \bibfield{author}{%
  \bibinfo {author} {\bibfnamefont{Y.}~\bibnamefont{Pang}}, \bibinfo {author}
  {\bibfnamefont{D.~E.}\ \bibnamefont{Kahana}},\ and\ \bibinfo {author}
  {\bibfnamefont{S.~H.}\ \bibnamefont{Kahana}},\ }%
  \bibfield{journal}{%
  \Doi{10.1103/PhysRevLett.78.3418}{\bibinfo {journal} {Phys. Rev. Lett.}}\ }%
  \textbf{\bibinfo {volume} {78}},\ \bibinfo {pages} {3418} (\bibinfo {year}
  {1997})%
  \bibAnnoteFile{NoStop}{Pang:1996ez}%
%%CITATION = NUCL-TH/9608014;%%
\bibitem{Spieles:1995fs}%
  \BibitemOpen
  \bibfield{author}{%
  \bibinfo {author} {\bibfnamefont{C.}~\bibnamefont{Spieles}}, \bibinfo
  {author} {\bibfnamefont{M.}~\bibnamefont{Bleicher}}, \bibinfo {author}
  {\bibfnamefont{A.}~\bibnamefont{Jahns}}, \bibinfo {author}
  {\bibfnamefont{R.}~\bibnamefont{Matiello}}, \bibinfo {author}
  {\bibfnamefont{H.}~\bibnamefont{Sorge}}, \bibinfo {author}
  {\bibfnamefont{H.}~\bibnamefont{St\"ocker}},\ and\ \bibinfo {author}
  {\bibfnamefont{W.}~\bibnamefont{Greiner}},\ }%
  \bibfield{journal}{%
  \Doi{10.1103/PhysRevC.53.2011}{\bibinfo {journal} {Phys. Rev.}}\ }%
  \textbf{\bibinfo {volume} {C53}},\ \bibinfo {pages} {2011} (\bibinfo {year}
  {1996})%
  \bibAnnoteFile{NoStop}{Spieles:1995fs}%
%%CITATION = NUCL-TH/9506008;%%
\bibitem{Pandharipande:1992zz}%
  \BibitemOpen
  \bibfield{author}{%
  \bibinfo {author} {\bibfnamefont{V.~R.}\ \bibnamefont{Pandharipande}}\ and\
  \bibinfo {author} {\bibfnamefont{S.~C.}\ \bibnamefont{Pieper}},\ }%
  \bibfield{journal}{%
  \Doi{10.1103/PhysRevC.45.791}{\bibinfo {journal} {Phys. Rev.}}\ }%
  \textbf{\bibinfo {volume} {C45}},\ \bibinfo {pages} {791} (\bibinfo {year}
  {1992})%
  \bibAnnoteFile{NoStop}{Pandharipande:1992zz}%
%%CITATION = PHRVA,C45,791;%%
\bibitem{Fuchs:2001fp}%
  \BibitemOpen
  \bibfield{author}{%
  \bibinfo {author} {\bibfnamefont{C.}~\bibnamefont{Fuchs}}, \bibinfo {author}
  {\bibfnamefont{A.}~\bibnamefont{Faessler}},\ and\ \bibinfo {author}
  {\bibfnamefont{M.}~\bibnamefont{El-Shabshiry}},\ }%
  \bibfield{journal}{%
  \Doi{10.1103/PhysRevC.64.024003}{\bibinfo {journal} {Phys. Rev.}}\ }%
  \textbf{\bibinfo {volume} {C64}},\ \bibinfo {pages} {024003} (\bibinfo {year}
  {2001})%
  \bibAnnoteFile{NoStop}{Fuchs:2001fp}%
%%CITATION = NUCL-TH/0103057;%%
\bibitem{TerHaar:1987ce}%
  \BibitemOpen
  \bibfield{author}{%
  \bibinfo {author} {\bibfnamefont{B.}~\bibnamefont{Ter~Haar}}\ and\ \bibinfo
  {author} {\bibfnamefont{R.}~\bibnamefont{Malfliet}},\ }%
  \bibfield{journal}{%
  \Doi{10.1103/PhysRevC.36.1611}{\bibinfo {journal} {Phys. Rev.}}\ }%
  \textbf{\bibinfo {volume} {C36}},\ \bibinfo {pages} {1611} (\bibinfo {year}
  {1987})%
  \bibAnnoteFile{NoStop}{TerHaar:1987ce}%
%%CITATION = PHRVA,C36,1611;%%
\bibitem{Larionov:2003av}%
  \BibitemOpen
  \bibfield{author}{%
  \bibinfo {author} {\bibfnamefont{A.~B.}\ \bibnamefont{Larionov}}\ and\
  \bibinfo {author} {\bibfnamefont{U.}~\bibnamefont{Mosel}},\ }%
  \bibfield{journal}{%
  \Doi{10.1016/j.nuclphysa.2003.08.005}{\bibinfo {journal} {Nucl. Phys.}}\ }%
  \textbf{\bibinfo {volume} {A728}},\ \bibinfo {pages} {135} (\bibinfo {year}
  {2003})%
  \bibAnnoteFile{NoStop}{Larionov:2003av}%
%%CITATION = NUCL-TH/0307035;%%
\bibitem{Pontecorvo:1956vx}%
  \BibitemOpen
  \bibfield{author}{%
  \bibinfo {author} {\bibfnamefont{B.~M.}\ \bibnamefont{Pontecorvo}},\ }%
  \bibfield{journal}{%
  \bibinfo {journal} {Zh. Eksp. Teor. Fiz.}\ }%
  \textbf{\bibinfo {volume} {30}},\ \bibinfo {pages} {947} (\bibinfo {year}
  {1956})%
  \bibAnnoteFile{NoStop}{Pontecorvo:1956vx}%
%%CITATION = ZETFA,30,947;%%
\bibitem{Minor:1992}%
  \BibitemOpen
  \bibfield{author}{%
  \bibinfo {author} {\bibfnamefont{E.~D.}\ \bibnamefont{Minor}}, \bibinfo
  {author} {\bibfnamefont{T.~A.}\ \bibnamefont{Armstrong}}, \bibinfo {author}
  {\bibfnamefont{B.}~\bibnamefont{Chen}}, \bibinfo {author}
  {\bibfnamefont{R.~A.}\ \bibnamefont{Lewis}},\ and\ \bibinfo {author}
  {\bibfnamefont{G.~A.}\ \bibnamefont{Smith}},\ }%
  \bibfield{journal}{%
  \bibinfo {journal} {Z. Phys.}\ }%
  \textbf{\bibinfo {volume} {A342}},\ \bibinfo {pages} {447} (\bibinfo {year}
  {1992})%
  \bibAnnoteFile{NoStop}{Minor:1992}%
\bibitem{Montagna:2002wg}%
  \BibitemOpen
  \bibfield{author}{%
  \bibinfo {author} {\bibfnamefont{P.}~\bibnamefont{Montagna}} \emph{et~al.}
  (\bibinfo {collaboration} {Obelix}),\ }%
  \bibfield{journal}{%
  \Doi{10.1016/S0375-9474(01)01302-1}{\bibinfo {journal} {Nucl. Phys.}}\ }%
  \textbf{\bibinfo {volume} {A700}},\ \bibinfo {pages} {159} (\bibinfo {year}
  {2002})%
  \bibAnnoteFile{NoStop}{Montagna:2002wg}%
%%CITATION = NUPHA,A700,159;%%
\bibitem{Mcgaughey:1986kz}%
  \BibitemOpen
  \bibfield{author}{%
  \bibinfo {author} {\bibfnamefont{P.~L.}\ \bibnamefont{McGaughey}}
  \emph{et~al.},\ }%
  \bibfield{journal}{%
  \Doi{10.1103/PhysRevLett.56.2156}{\bibinfo {journal} {Phys. Rev. Lett.}}\ }%
  \textbf{\bibinfo {volume} {56}},\ \bibinfo {pages} {2156} (\bibinfo {year}
  {1986})%
  \bibAnnoteFile{NoStop}{Mcgaughey:1986kz}%
%%CITATION = PRLTA,56,2156;%%
\bibitem{Cugnon:1989mx}%
  \BibitemOpen
  \bibfield{author}{%
  \bibinfo {author} {\bibfnamefont{J.}~\bibnamefont{Cugnon}}, \bibinfo {author}
  {\bibfnamefont{P.}~\bibnamefont{Deneye}},\ and\ \bibinfo {author}
  {\bibfnamefont{J.}~\bibnamefont{Vandermeulen}},\ }%
  \bibfield{journal}{%
  \Doi{10.1016/0375-9474(89)90236-4}{\bibinfo {journal} {Nucl. Phys.}}\ }%
  \textbf{\bibinfo {volume} {A500}},\ \bibinfo {pages} {701} (\bibinfo {year}
  {1989})%
  \bibAnnoteFile{NoStop}{Cugnon:1989mx}%
%%CITATION = NUPHA,A500,701;%%
\bibitem{Moussallam:1984uj}%
  \BibitemOpen
  \bibfield{author}{%
  \bibinfo {author} {\bibfnamefont{B.}~\bibnamefont{Moussallam}},\ }%
  \bibfield{journal}{%
  \Doi{10.1016/0375-9474(83)90659-0}{\bibinfo {journal} {Nucl. Phys.}}\ }%
  \textbf{\bibinfo {volume} {A407}},\ \bibinfo {pages} {413} (\bibinfo {year}
  {1983})%
  \bibAnnoteFile{NoStop}{Moussallam:1984uj}%
%%CITATION = NUPHA,A407,413;%%
\bibitem{Moussallam:1984zj}%
  \BibitemOpen
  \bibfield{author}{%
  \bibinfo {author} {\bibfnamefont{B.}~\bibnamefont{Moussallam}},\ }%
  \bibfield{journal}{%
  \Doi{10.1016/0375-9474(84)90690-0}{\bibinfo {journal} {Nucl. Phys.}}\ }%
  \textbf{\bibinfo {volume} {A429}},\ \bibinfo {pages} {429} (\bibinfo {year}
  {1984})%
  \bibAnnoteFile{NoStop}{Moussallam:1984zj}%
%%CITATION = NUPHA,A429,429;%%
\bibitem{Eastman:1973va}%
  \BibitemOpen
  \bibfield{author}{%
  \bibinfo {author} {\bibfnamefont{P.~S.}\ \bibnamefont{Eastman}}
  \emph{et~al.},\ }%
  \bibfield{journal}{%
  \Doi{10.1016/0550-3213(73)90499-9}{\bibinfo {journal} {Nucl. Phys.}}\ }%
  \textbf{\bibinfo {volume} {B51}},\ \bibinfo {pages} {29} (\bibinfo {year}
  {1973})%
  \bibAnnoteFile{NoStop}{Eastman:1973va}%
%%CITATION = NUPHA,B51,29;%%
\bibitem{Tanimori:1985ps}%
  \BibitemOpen
  \bibfield{author}{%
  \bibinfo {author} {\bibfnamefont{T.}~\bibnamefont{Tanimori}} \emph{et~al.},\
  }%
  \bibfield{journal}{%
  \Doi{10.1103/PhysRevLett.55.1835}{\bibinfo {journal} {Phys. Rev. Lett.}}\ }%
  \textbf{\bibinfo {volume} {55}},\ \bibinfo {pages} {1835} (\bibinfo {year}
  {1985})%
  \bibAnnoteFile{NoStop}{Tanimori:1985ps}%
%%CITATION = PRLTA,55,1835;%%
\bibitem{Bardin:1987hw}%
  \BibitemOpen
  \bibfield{author}{%
  \bibinfo {author} {\bibfnamefont{G.}~\bibnamefont{Bardin}} \emph{et~al.},\ }%
  \bibfield{journal}{%
  \Doi{10.1016/0370-2693(87)90140-7}{\bibinfo {journal} {Phys. Lett.}}\ }%
  \textbf{\bibinfo {volume} {B192}},\ \bibinfo {pages} {471} (\bibinfo {year}
  {1987})%
  \bibAnnoteFile{NoStop}{Bardin:1987hw}%
%%CITATION = PHLTA,B192,471;%%
\bibitem{Sugimoto:1987ct}%
  \BibitemOpen
  \bibfield{author}{%
  \bibinfo {author} {\bibfnamefont{Y.}~\bibnamefont{Sugimoto}} \emph{et~al.},\
  }%
  \bibfield{journal}{%
  \Doi{10.1103/PhysRevD.37.583}{\bibinfo {journal} {Phys. Rev.}}\ }%
  \textbf{\bibinfo {volume} {D37}},\ \bibinfo {pages} {583} (\bibinfo {year}
  {1988})%
  \bibAnnoteFile{NoStop}{Sugimoto:1987ct}%
%%CITATION = PHRVA,D37,583;%%
\bibitem{Bardin:1994am}%
  \BibitemOpen
  \bibfield{author}{%
  \bibinfo {author} {\bibfnamefont{G.}~\bibnamefont{Bardin}} \emph{et~al.},\ }%
  \bibfield{journal}{%
  \Doi{10.1016/0550-3213(94)90052-3}{\bibinfo {journal} {Nucl. Phys.}}\ }%
  \textbf{\bibinfo {volume} {B411}},\ \bibinfo {pages} {3} (\bibinfo {year}
  {1994})%
  \bibAnnoteFile{NoStop}{Bardin:1994am}%
%%CITATION = NUPHA,B411,3;%%
\bibitem{Kim:1996ada}%
  \BibitemOpen
  \bibfield{author}{%
  \bibinfo {author} {\bibfnamefont{H.}~\bibnamefont{Kim}}, \bibinfo {author}
  {\bibfnamefont{S.}~\bibnamefont{Schramm}},\ and\ \bibinfo {author}
  {\bibfnamefont{S.~H.}\ \bibnamefont{Lee}},\ }%
  \bibfield{journal}{%
  \Doi{10.1103/PhysRevC.56.1582}{\bibinfo {journal} {Phys. Rev.}}\ }%
  \textbf{\bibinfo {volume} {C56}},\ \bibinfo {pages} {1582} (\bibinfo {year}
  {1997})%
  \bibAnnoteFile{NoStop}{Kim:1996ada}%
%%CITATION = NUCL-TH/9610011;%%
\bibitem{Wagner:2004ee}%
  \BibitemOpen
  \bibfield{author}{%
  \bibinfo {author} {\bibfnamefont{M.}~\bibnamefont{Wagner}}, \bibinfo {author}
  {\bibfnamefont{A.~B.}\ \bibnamefont{Larionov}},\ and\ \bibinfo {author}
  {\bibfnamefont{U.}~\bibnamefont{Mosel}},\ }%
  \bibfield{journal}{%
  \Doi{10.1103/PhysRevC.71.034910}{\bibinfo {journal} {Phys. Rev.}}\ }%
  \textbf{\bibinfo {volume} {C71}},\ \bibinfo {pages} {034910} (\bibinfo {year}
  {2005})%
  \bibAnnoteFile{NoStop}{Wagner:2004ee}%
%%CITATION = NUCL-TH/0411010;%%
\bibitem{Golubeva:1992tr}%
  \BibitemOpen
  \bibfield{author}{%
  \bibinfo {author} {\bibfnamefont{E.~S.}\ \bibnamefont{Golubeva}}, \bibinfo
  {author} {\bibfnamefont{A.~S.}\ \bibnamefont{Iljinov}}, \bibinfo {author}
  {\bibfnamefont{B.~V.}\ \bibnamefont{Krippa}},\ and\ \bibinfo {author}
  {\bibfnamefont{I.~A.}\ \bibnamefont{Pshenichnov}},\ }%
  \bibfield{journal}{%
  \Doi{10.1016/0375-9474(92)90362-N}{\bibinfo {journal} {Nucl. Phys.}}\ }%
  \textbf{\bibinfo {volume} {A537}},\ \bibinfo {pages} {393} (\bibinfo {year}
  {1992})%
  \bibAnnoteFile{NoStop}{Golubeva:1992tr}%
%%CITATION = NUPHA,A537,393;%%
\bibitem{PshenichnovPhD}%
  \BibitemOpen
  \bibfield{author}{%
  \bibinfo {author} {\bibfnamefont{I.~A.}\ \bibnamefont{Pshenichnov}},\ }%
  \bibinfo {note} {doctoral thesis, INR, Moscow, 1998}%
  \bibAnnoteFile{NoStop}{PshenichnovPhD}%
\bibitem{Friedman:2007zz}%
  \BibitemOpen
  \bibfield{author}{%
  \bibinfo {author} {\bibfnamefont{E.}~\bibnamefont{Friedman}}\ and\ \bibinfo
  {author} {\bibfnamefont{A.}~\bibnamefont{Gal}},\ }%
  \bibfield{journal}{%
  \Doi{10.1016/j.physrep.2007.08.002}{\bibinfo {journal} {Phys. Rept.}}\ }%
  \textbf{\bibinfo {volume} {452}},\ \bibinfo {pages} {89} (\bibinfo {year}
  {2007})%
  \bibAnnoteFile{NoStop}{Friedman:2007zz}%
%%CITATION = 0705.3965;%%
\bibitem{Seki:1983sh}%
  \BibitemOpen
  \bibfield{author}{%
  \bibinfo {author} {\bibfnamefont{R.}~\bibnamefont{Seki}}\ and\ \bibinfo
  {author} {\bibfnamefont{K.}~\bibnamefont{Masutani}},\ }%
  \bibfield{journal}{%
  \Doi{10.1103/PhysRevC.27.2799}{\bibinfo {journal} {Phys. Rev.}}\ }%
  \textbf{\bibinfo {volume} {C27}},\ \bibinfo {pages} {2799} (\bibinfo {year}
  {1983})%
  \bibAnnoteFile{NoStop}{Seki:1983sh}%
%%CITATION = PHRVA,C27,2799;%%
\bibitem{Friedman:1983wz}%
  \BibitemOpen
  \bibfield{author}{%
  \bibinfo {author} {\bibfnamefont{E.}~\bibnamefont{Friedman}},\ }%
  \bibfield{journal}{%
  \Doi{10.1103/PhysRevC.28.1264}{\bibinfo {journal} {Phys. Rev.}}\ }%
  \textbf{\bibinfo {volume} {C28}},\ \bibinfo {pages} {1264} (\bibinfo {year}
  {1983})%
  \bibAnnoteFile{NoStop}{Friedman:1983wz}%
%%CITATION = PHRVA,C28,1264;%%
\bibitem{Dalkarov:1986tu}%
  \BibitemOpen
  \bibfield{author}{%
  \bibinfo {author} {\bibfnamefont{O.~D.}\ \bibnamefont{Dalkarov}}\ and\
  \bibinfo {author} {\bibfnamefont{V.~A.}\ \bibnamefont{Karmanov}},\ }%
  \bibfield{journal}{%
  \Doi{10.1016/0375-9474(85)90561-5}{\bibinfo {journal} {Nucl. Phys.}}\ }%
  \textbf{\bibinfo {volume} {A445}},\ \bibinfo {pages} {579} (\bibinfo {year}
  {1985})%
  \bibAnnoteFile{NoStop}{Dalkarov:1986tu}%
%%CITATION = NUPHA,A445,579;%%
\bibitem{Janouin:1986vh}%
  \BibitemOpen
  \bibfield{author}{%
  \bibinfo {author} {\bibfnamefont{S.}~\bibnamefont{Janouin}} \emph{et~al.},\
  }%
  \bibfield{journal}{%
  \Doi{10.1016/0375-9474(86)90291-5}{\bibinfo {journal} {Nucl. Phys.}}\ }%
  \textbf{\bibinfo {volume} {A451}},\ \bibinfo {pages} {541} (\bibinfo {year}
  {1986})%
  \bibAnnoteFile{NoStop}{Janouin:1986vh}%
%%CITATION = NUPHA,A451,541;%%
\bibitem{Friedman:1986nf}%
  \BibitemOpen
  \bibfield{author}{%
  \bibinfo {author} {\bibfnamefont{E.}~\bibnamefont{Friedman}}\ and\ \bibinfo
  {author} {\bibfnamefont{J.}~\bibnamefont{Lichtenstadt}},\ }%
  \bibfield{journal}{%
  \Doi{10.1016/0375-9474(86)90451-3}{\bibinfo {journal} {Nucl. Phys.}}\ }%
  \textbf{\bibinfo {volume} {A455}},\ \bibinfo {pages} {573} (\bibinfo {year}
  {1986})%
  \bibAnnoteFile{NoStop}{Friedman:1986nf}%
%%CITATION = NUPHA,A455,573;%%
\bibitem{Wong:1984fy}%
  \BibitemOpen
  \bibfield{author}{%
  \bibinfo {author} {\bibfnamefont{C.~Y.}\ \bibnamefont{Wong}}, \bibinfo
  {author} {\bibfnamefont{A.~K.}\ \bibnamefont{Kerman}}, \bibinfo {author}
  {\bibfnamefont{G.~R.}\ \bibnamefont{Satchler}},\ and\ \bibinfo {author}
  {\bibfnamefont{A.~D.}\ \bibnamefont{Mackellar}},\ }%
  \bibfield{journal}{%
  \Doi{10.1103/PhysRevC.29.574}{\bibinfo {journal} {Phys. Rev.}}\ }%
  \textbf{\bibinfo {volume} {C29}},\ \bibinfo {pages} {574} (\bibinfo {year}
  {1984})%
  \bibAnnoteFile{NoStop}{Wong:1984fy}%
%%CITATION = PHRVA,C29,574;%%
\end{thebibliography}%

\end{document}